\documentclass{aa}
\usepackage{graphicx}
\newcommand{\simleq}{\; \raisebox{-0.4ex}{\tiny$\stackrel{{\textstyle<}}{\sim}$}\;} 
\newcommand{\simgeq}{\; \raisebox{-0.4ex}{\tiny$\stackrel{{\textstyle>}}{\sim}$}\;} 
\newcommand{\new}[1]{{{#1}}}
\newcommand{\newnew}[1]{{{#1}}}
\begin{document}
   \title{The co-orbital corotation torque in a viscous disk: numerical
simulations}

   \author{F.S. Masset
          }

   \offprints{F.S. Masset}

   \institute{Service d'Astrophysique, L'Orme des Merisiers,
              CE-Saclay, 91191 Gif/Yvette Cedex, France\\
              \email{fmasset@cea.fr}
             }

   \date{Received; accepted}

   \abstract{The torque felt by a non-accreting protoplanet on a 
circular orbit embedded in a uniform surface density protoplanetary disk is analyzed by 
means of time-dependent numerical 
simulations. Varying the viscosity enables one to disentangle the Lindblad
torque (which is independent of viscosity) from the corotation torque,
which saturates at low viscosity and is unsaturated at high viscosity.
The dependence of the corotation torque upon the viscosity and upon the
width of the librating zone is compared with 
previous analytical expressions, and shown to be in agreement with 
those. The effect of the potential smoothing respectively on the Lindblad
torque and on the corotation torque is investigated, and the question of whether
3D effects and their impact on the total torque sign and magnitude can be
modeled by an adequate smoothing prescription in a 2D simulation is addressed.
As a side result, this study shows that the total torque acting on a Neptune-sized protoplanet
is positive in a
sufficiently thin, 
viscous disk ($H/r \simleq 4$\%, $\alpha \simgeq 10^{-2}$), but the inward migration time of smaller
bodies is still \new{very} short, making it unlikely that they reach the torque reversal mass before
having migrated all the way to the central object.
   \keywords{planet formation --
        protoplanetary disks -- tidal interactions -- hydrodynamics -- methods: Numerical
               }
   }

   \maketitle
%

\section{Introduction}
\label{sec:intro}
The formation of planets is likely to occur in the protoplanetary disks surrounding
the very young stars. One of the most widely accepted formation scenario is the core accretion one,
whereby a solid core is slowly built up by the accretion of solid material (rocks and ice) and
subsequently, if its mass becomes large enough, by the rapid accretion of the surrounding gas.
As the core builds up, it gravitationally interacts with the surrounding nebula, and as a result of
the exchange of angular momentum and energy with the nebula, it undergoes a change
in its orbital parameters. The basis of the gravitational interaction between a disk and an orbiting
point-like perturber was worked out about twenty years ago (Goldreich \& Tremaine 1979 and 1980,
Lin \& Papaloizou 1979) and since then refined in great detail (Ward 1986 and 1988, Artymowicz 1993,
Ward 1997). The most striking prediction of these works was that a protoplanet embedded in a 
protoplanetary disk would undergo a rapid orbital decay or inward migration, and this expectation
has received strong observational support from the discovery of short-period giant planets
(Mayor \& Queloz 1995), the formation of which is unlikely to have occurred {\it in situ}. Most
of the work done on the planet-disk tidal interaction has been done in the linear regime. 
Each azimuthal Fourier component of the planet perturbing potential can be shown to contribute sizably to the disk-planet
tidal interaction wherever its frequency in the local frame either vanishes or matches $\pm\kappa$, the
epicyclic frequency (which coincides with the Keplerian frequency in a Keplerian disk such as a protoplanetary
disk, where self-gravity, at least in the central parts, in considered as unimportant). The angular momentum
exchange rate through the terms for which the perturber frequency in the local frame is $+\kappa$ (resp.
$-\kappa$) is called the outer Lindblad torque (resp. the inner Lindblad torque), while the terms corresponding
to a vanishing perturbed frequency in the local frame correspond to the corotation torque. If one makes
the assumption of a circular orbit (which is reasonable as it can be shown that the global effect
of the resonances is an eccentricity damping in most cases, see e.g. Papaloizou \& al. 2001 and references
therein), then the Outer Lindblad Resonances (OLR)
fall exterior to the orbit, the Inner Lindblad Resonances (ILR) fall interior to it, and the corotation resonances
nearly coincide with the orbit (they actually lie slightly inside of this latter as the disk is slightly sub-Keplerian
due to the partial support of gravity by a radial pressure gradient), and the corresponding torque is called for obvious
reasons the co-orbital corotation torque. The outer Lindblad torque can be shown to
be negative, while the inner Lindblad torque is positive. There exists a mismatch between both (almost always
in favor of the outer Lindblad torque, Ward 1997) which scales as~$h$, the disk aspect ratio, and which leads
to the inward migration. The Lindblad torque has been shown to be independent of viscosity (Meyer-Vernet
\& Sicardy 1987, and Papaloizou \& Lin 1984). \newnew{This is true as long as the surface
density profile itself is not sizably modified under the action of the one-sided Lindblad torque, i.e. as long as the relative
depth of the dip opened around the orbit by the planet is small.
This assumption is fulfilled in the present work except for the highest masses and the lowest
viscosities (see also discussion at section~\ref{sec:dippb})}. On the other hand, the corotation torque in the linear regime
does not correspond to an angular momentum flux carried away by waves. The angular momentum rather
accumulates at corotation, and the corresponding torque appears as a discontinuity in the angular momentum
flux at corotation (Goldreich \& Tremaine 1979). Ward (1992) has shown, by summing the corotation
resonances after an adequate vertical averaging of the perturbing potential, that the corotation torque is
at most comparable to the differential Lindblad torque, whereas Korycansky \& Pollack (1993) have shown
by solving numerically the linearized equations for a steady state perturbation of the disk that the actual solution
was even smaller than the analytical estimates. The corotation torque scales as the gradient of specific vorticity across
corotation $\partial(\Sigma/B)/\partial r$, where $B$ is the second Oort's constant and is $\Omega/4$ in a
Keplerian disk. Ward (1991) has shown that the corotation torque comes from the exchange of angular momentum
between the planet and the near-by fluid elements which move from one side to the other of their horseshoe streamline,
and that the dependency upon the specific vorticity gradient was due to an uneven mapping of the fluid elements
between the outer and inner leg of their horseshoe streamline. In the linear regime, i.e. as the planet mass tends to zero, 
the horseshoe zone width tends to zero, while the libration time of the fluid elements on the horseshoe streamlines tends to infinity,
which is in agreement with the picture of a localized and constant discontinuity of the angular momentum flux at corotation.
In the finite mass regime the situation is quite different however. The libration time can be much shorter than the migration
time across the horseshoe region, and also much shorter than the viscous diffusion time, in which case libration removes
the specific vorticity gradient (Ward 1992), and the corotation torque therefore vanishes (saturates) after a few libration times. In a sufficiently
viscous disk however, the viscous diffusion and the radial transport of material across the horseshoe region may inhibit
the  corotation torque saturation. A way of evaluating the co-orbital corotation torque as a function of viscosity has been given
by Masset (2001), for a planet held on a circular orbit in a uniform surface density disk, and for a steady flow in the planet frame. 
As the material set in libration in a steady flow is trapped, the total torque
on the librating fluid elements region vanishes. This torque is the sum of the viscous torque on the trapped region and the
gravitational torque of the planet, which is opposite to one component of the corotation torque. As the viscous force is a contact
action, the integral of the viscous torque on the trapped region reduces to the torque exerted on its boundary (the separatrix
between circulating and librating streamlines) by the outer and inner disks. The second and last component of the corotation
torque is given by the angular momentum lost by the fluid elements which participate in the global accretion of the disk
material onto the central object (accretion of which are excluded only the fluid elements of the librating region), when they flow from
the outer to the inner disk and undergo one horseshoe like close encounter with the planet. The corotation torque can therefore
be expressed (in the steady state case) only with the knowledge of the flow properties at the separatrix. The flow properties in
the librating region can be split into  an even and an odd part of the perturbed density (w.r.t the distance to corotation). The former
comes from the perturbations of the surface density and azimuthal velocity profiles under the action of the one-sided Lindblad
torque, while the latter comes from 
libration\footnote{\newnew{Libration will tend indeed to flatten radially across the horseshoe region
 the profile of any quantity conserved along a streamline. This is the case
of the specific vorticity $B/\Sigma$ in an inviscid disk. Libration therefore tends to impose a decreasing profile of $\Sigma$ around
corotation in an inviscid disk. Only in the special case of the shearing sheet, in which one neglects the radial variation of $B$, does
libration tend to eliminate in-out surface density differences.}}.
Masset (2001) gives the following expression for the corotation torque:
\begin{equation}
\label{eq:finale}
\Gamma_C=\Gamma_M^C+\Gamma_A^C
\end{equation}
where $\Gamma_M^C$ is the main term, which comes from the odd part of the perturbed density and which therefore is linked
to libration and exhibits a dependence on the viscosity:
\begin{equation}
\label{eq:main}
\Gamma_M^C=\frac 98x_s^4\Omega_p^2\Sigma_0{\cal F}(z_s)
\end{equation}
where ${\cal F}(z_s)$ is a function of the viscosity and the horseshoe zone width $x_s$ (in which, contrary to Masset 2001,
a factor~4 has been introduced for convenience, {\it cf. infra}),
while $\Gamma_A^C$ is an additional term which comes from the even part of the perturbed density, and which introduces a coupling
between the one-sided Lindblad torque and the corotation torque:
\begin{equation}
\label{eq:add}
\Gamma_A^C=\frac{x_s}{r_p}{\cal G}(x_s)\Gamma_{LR}
\end{equation}
where ${\cal G}(x_s)=O(1)$ is a function which depends on the exact position of the separatrix, and where $\Gamma_\mathrm{LR}$
is the one-sided Lindblad torque (the arithmetical average of the Outer and Inner Lindblad torques absolute values). The fact that this additional
term does not exhibit any dependence on the viscosity may seem paradoxical, as one expects the co-orbital corotation torque to
vanish in an inviscid disk. This paradox is only apparent however. This term comes from the even terms in the perturbed
density (and odd in the perturbed azimuthal velocity), which scale as $\nu^{-1}$, therefore this dependence cancels out with the
$\nu$ overall dependence of the viscous torque on the separatrix. This results holds as long as the even perturbed density does not
saturate (i.e. as long as the dip is shallow enough and does not correspond to a gap). In the strictly inviscid limit (and provided
that in this limit one can have a steady state situation), the planet (which in this
artificial situation is held on a fixed circular orbit) opens a gap, and therefore
the even perturbed density term saturates and the additional term given by Eq.~(\ref{eq:add}) vanishes.

The purpose of this paper is to investigate this situation and to check by means of numerical simulations the validity of Eqs.~(\ref{eq:main})
and~(\ref{eq:add}). For that purpose, numerical simulations are performed over a large range of
viscosities. 
Other questions addressed in this paper are the ability of 2D numerical codes to predict correctly the total torque
acting  on an embedded point-like object through the choice of a correct value  of the potential smoothing length, and the possibility
of migration reversal (i.e. outward) due to the corotation torque, although linear regime studies seem to conclude that the
(positive) corotation torque is too weak to counteract the (negative) differential Lindblad torque. 
Although the additional term of Eq.~(\ref{eq:add}) is too small to cancel out the differential Lindblad torque by itself, it is of
interest to examine the situation for weakly embedded objects where the corotation torque is
lift up by the additional term of Eq.~(\ref{eq:add}), and where the differential Lindblad torque
may be reduced with respect to its linear value (see e.g. Miyoshi et al. 1999, Lin \& Papaloizou 1979).

It should be noted that global simulations are needed to describe properly the corotation torque. Any local description, such as the one
provided by a box or a wedge centered on the planet, will prevent saturation if inflow/outflow boundary conditions are used in
azimuth, which inject ``fresh'' material  with an unperturbed density and velocity profile, while the use of periodic boundary conditions
in these configurations artificially shortens the libration time. Furthermore, the use of the shearing sheet formalism, which usually neglects
the radial derivative of the unperturbed disk specific vorticity, leads to a dependence of the corotation torque on the radial derivative
of the surface density, thus switching off the corotation torque in the convenient situation where the surface density is uniform.

\section{Notations}

The planet orbital radius is $r_p$, and its orbital frequency $\Omega_p$. The position
of a fluid element in the disk is represented by its polar coordinates radius $r$ and azimuth $\theta$,
counted rotation-wise, with its origin in the planet direction. The distance of a fluid element
to the planet \new{orbit} is $x=r-r_p$, and its dimensionless counterpart is $\hat x= x/r_p$. The Keplerian frequency
is $\Omega_K(r)$, and the disk material orbital frequency is $\Omega(r)$.
The disk kinematic viscosity
is $\nu$ and its dimensionless counterpart is $\hat\nu=\nu/(r_p^2\Omega_p)$. The $\alpha$ parameterization
(Shakura \& Syunyaev 1973) will also be used, with $\nu=\alpha Hc_s$, $H$ being the disk vertical
scale length and $c_s$ being the sound speed.
The disk surface density
is denoted with $\Sigma$, while its unperturbed uniform value is with $\Sigma_0$. The planet mass is $m_p$, the
central object mass is $M_*$, and the ratio of both is $q=m_p/M_*$. The torque exerted on the planet
is denoted $\Gamma$, and its non-dimensional counterpart is $\hat\Gamma=\Gamma/\Gamma_0$
where $\Gamma_0=\pi r_p^4\Omega_p^2q^2\Sigma_0/h^3$, $h$ being the aspect ratio. The perturbed
velocity has radial and azimuthal components $v_r$ and $v_\theta$.

\section{Numerical aspects}
\label{sec:numeraspects}
The code that was used is an Eulerian ZEUS-like 2D code based on a polar
grid centered on the primary (Stone \& Norman 1992, Nelson et al. 2000),
and corotating with the planet (Kley 1998).
The runs described here were performed using a modified azimuthal
Courant condition (Masset 2000) in order to increase the timestep, and some 
of these runs were checked against the standard azimuthal transport procedure,
and found to give almost identical results. The grid consists of 
$N_\theta=450$ by $N_r=143$ zones, uniformly spaced in azimuth and radius.
The planet lies at radius $r_p=1$ and azimuth $\theta=0$. The central star mass
and gravitational constant are respectively $M_*=1$ and $G=1$, and the
time unit is chosen to be $(r_p/GM_*)^{3/2}=\Omega_p^{-1}$, so that
the planet orbital period in our unit system is $2\pi$. The grid outer boundary
is chosen to be at $r=2.5$, while the inner boundary is at $r=0.504651$,
so that the planet is located just at the center of a zone, in order to 
limit any bias in the torque and in the co-orbital dynamics due to an uneven
placement of the planet w.r.t. the grid. 

\subsection{Basic equations}
The equations solved by 
the code are:

\begin{itemize}
\item The continuity equation, which reads:

\begin{equation}
\frac{\partial\Sigma}{\partial t}+\frac 1r\frac{\partial(rv_r\Sigma)}{\partial
r}+\frac 1r\frac{\partial(v_\theta\Sigma)}{\partial\theta}=0
\end{equation}

\item The Navier-Stokes equations, which reads respectively for $v_r$ and
$v_\theta$:

\begin{equation}
\frac{\partial v_r}{\partial t}
+v_r\frac{\partial v_r}{\partial r}
+\frac{v_\theta}{r}\frac{\partial v_r}{\partial \theta}
-\frac{v_\theta^2}{r}=-\frac 1\Sigma\frac{\partial p}{\partial r}
-\frac{\partial\phi}{\partial r}
+\frac{f_r}\Sigma
\end{equation}

and~:

\begin{equation}
\frac{\partial v_\theta}{\partial t}
+v_r\frac{\partial v_\theta}{\partial r}
+\frac{v_\theta}{r}\frac{\partial v_\theta}{\partial \theta}
+\frac{v_rv_\theta}{r}=-\frac 1{\Sigma r}\frac{\partial p}{\partial \theta}
-\frac 1r\frac{\partial\phi}{\partial \theta}
+\frac{f_\theta}\Sigma
\end{equation}

where $p$ is the vertically integrated pressure, $f_r$ and $f_\theta$ respectively the radial and azimuthal component of the viscous force per unit surface, and $\phi$ is
the gravitational potential.

\item The equation of state is that of an isothermal gas with
 sound speed
$c_s$:
\begin{equation}
p=c_s^2\Sigma
\end{equation}

\item The gravitational potential includes the potential of the primary,
the potential of the protoplanet, and an indirect term which
arises from the fact that the non-rotating frame centered on the primary is not inertial:
\begin{eqnarray}
\label{eq:indirect}
\phi(r,\theta)&=&-\frac{GM_*}{r}-\frac{Gm_p}{(r^2+r_p^2-2rr_p\cos\theta+
\epsilon^2)^{1/2}} \nonumber \\ 
&&+\frac{Gm_p}{r_p^2}{r\cos\theta} \nonumber \\ 
&&+r\int_{\mathrm{disk}}\frac{G\Sigma(r',\theta')}{r'^2}
\cos(\theta-\theta')r'dr'd\theta'
\end{eqnarray}

In this expression, $\epsilon$ is the smoothing length of the protoplanet
potential, and otherwise stated it amounts to $60$~\% of the
disk ``thickness'' $H=c_s/\Omega$.

\item The viscous force is derived from the viscous stress tensor as follows:
\begin{eqnarray}
f_r &=& \frac{1}{r} \frac{\partial (r \tau_{rr})}{\partial r}  +
 \frac{1}{r} \frac{\partial \tau_{r \phi}}{\partial \phi} -
 \frac{\tau_{\phi \phi}}{r} \label{visc-r} \\
f_{\phi} &=& \frac{1}{r} \frac{\partial (r \tau_{\phi r})}{\partial r} +
 \frac{1}{r} \frac{\partial \tau_{\phi \phi}}{\partial \phi} +
 \frac{\tau_{r \phi}}{r}, \label{visc-phi}
\end{eqnarray}

where the components of the viscous stress tensor are:
\begin{eqnarray}
\tau_{rr} & = & 2 \eta D_{rr} - \frac{2}{3} \eta \nabla . {\bf v} \\ \nonumber
\tau_{\phi \phi} & = & 2 \eta D_{\phi \phi} - \frac{2}{3} \eta \nabla .
{\bf v}
\\ \nonumber
\tau_{r \phi} & = & \tau_{\phi r} = 2 \eta D_{r \phi}, \label{visc-tensor}
\end{eqnarray}
where
\begin{eqnarray}
D_{rr} &=& \frac{\partial v_r}{\partial r}, D_{\phi \phi} = \frac{1}{r}
\frac{\partial v_{\phi}}{\partial \phi} + \frac{v_r}{r} \\ \nonumber
D_{r \phi} &=& \frac{1}{2} \left[ r \frac{\partial}{\partial r} \left(
\frac{v_{\phi}}{r} \right) + \frac{1}{r} \frac{\partial v_r}{\partial \phi}
\right], \label{D}
\end{eqnarray}
and $\eta=\Sigma \nu$ is the vertically integrated
 dynamical viscosity coefficient.

\end{itemize}

\subsection{Boundary conditions}

The boundary conditions are $2\pi$-periodic in $\theta$ in order
to account for the disk geometry. 
As the viscosity can be large in some runs, it is important to take adequate
boundary conditions, otherwise the fast radial redistribution of disk material
can affect the slope of the surface density profile, which in turn can significantly
alter the corotation torque magnitude. For this reason, it is important :

\begin{itemize}
\item to have a source of material with surface density $\Sigma_0$ at large
radius in order to avoid outer disk depletion;
\item to allow the outflow (inwards) of disk material at the inner boundary only if
the disk surface density is larger or equal to $\Sigma_0$ (otherwise a positive
radial gradient of surface density develops, and one may overestimate the corotation
torque in these conditions).
\end{itemize}

\subsection{Initial setup}

In all the runs presented here, the unperturbed disk surface density is
uniform and is, in our unit system, $\Sigma_0=10^{-4}$. This translates
into:

\begin{equation}
\Sigma=\Sigma_0\cdot 8.9\cdot 10^6\cdot\left(\frac{M_*}{M_\odot}\right)
\cdot\left(\frac{r_p}{1\mbox{~AU}}\right)^{-2}\mbox{~g.cm}^{-2}
\end{equation}

Note that since the results presented here are scaled by $\Gamma_0$, they are
independent of the actual value of $\Sigma_0$ (except for the disk 
potential indirect term of Eq.~(\ref{eq:indirect}), which for the value chosen above is negligible).

The disk temperature profile is constant in time and is chosen such that
the disk aspect ratio~$h=H/r$ be uniform. The initial azimuthal velocity 
is computed
accordingly and is slightly sub-Keplerian due to the partial support of
gravity by the radial pressure gradient:

\begin{equation}
v_\theta=\left[\frac{GM_*}{r}(1-h^2)\right]^{1/2}
\end{equation}

The initial radial velocity is set to zero.

\subsection{Initial parameters}

A number of runs have been performed varying the planet mass, the disk aspect
ratio, the viscosity, and in some cases the smoothing prescription or
the resolution (which was then chosen twice higher). Each run consisted of
$120$~planet orbits (which was assumed to be sufficient to reach
a steady state in the planet frame). For a given choice of the planet
mass and disk aspect ratio, usually 13 runs were performed for different
values of the viscosity, logarithmicly spaced:
\begin{equation}
\hat\nu_i=10^{-7+2i/7}\mbox{~~~~~}(1\leq i\leq13)
\end{equation}

The set of main runs is described at Tab.~\ref{tab:mainruns}. The last cell of the $h=6$~\% line
is empty: the corresponding sets of runs have not been performed, as the corresponding
run would correspond to a horseshoe region radially not wider than one zone, and even
if one disregards finite resolution effects, the mismatch between corotation and orbit
is not negligible compared to the horseshoe zone half-width, which is one of the condition
of validity of Eqs.~(\ref{eq:main}) and~(\ref{eq:add}). \new{The first cell of the $h=3$~\% line
is also empty, as the situation it would describe would rather correspond to type~II migration, i.e.
the planet opens a gap around its orbit and the corotation torque is switched off, except for the largest
viscosities}.

   \begin{table}
        \centering
      \caption[]{The set of 247 main runs. A run with mass ratio~$q$, aspect ratio~$h$
and viscosity $\hat\nu_i$ is labeled R$m_h^i$, where $h$ is in percent and where $m$ 
is the round-off of the planet mass, in earth masses, if $M_*=M_\odot$. For instance,
run~R$6_3^8$ stands for the run with $q=1.67\cdot10^{-5}$, aspect ratio $h=0.03$, and
viscosity $\hat\nu_8$. The bulk of the runs are R$m_h^i$, with $m=2,6,11,17,30$, 
$h=3,4,5,6$~\% and $i=1$~to~$13$. Runs with $m=30$  have been performed for
extrapolation purposes only, since for such a mass a protoplanet is expected to rapidly
accrete the surrounding gas.}
         \label{tab:mainruns}
     $$ 
         \begin{array}{|l|cccc|}
            \hline
        _q\, ^{h (\%)} & 3 & 4 & 5 & 6\\
\hline
        6.67\cdot 10^{-6} & \nu_{1\rightarrow 13}& \nu_{1\rightarrow 13}& \nu_{1\rightarrow 13}& n/a\\
        1.67\cdot 10^{-5} & \nu_{1\rightarrow 13}& \nu_{1\rightarrow 13}& \nu_{1\rightarrow 13}& \nu_{1\rightarrow 13}\\
        3.33\cdot 10^{-5} & \nu_{1\rightarrow 13}& \nu_{1\rightarrow 13}& \nu_{1\rightarrow 13}& \nu_{1\rightarrow 13}\\
        5\cdot 10^{-5} & \nu_{1\rightarrow 13}& \nu_{1\rightarrow 13}& \nu_{1\rightarrow 13}& \nu_{1\rightarrow 13}\\
        10^{-4} & n/a& \nu_{1\rightarrow 13}& \nu_{1\rightarrow 13}& \nu_{1\rightarrow 13}\\
            \hline
         \end{array}
     $$ 
   \end{table}

Subsidiary runs have been performed in which either the smoothing or the resolution has been
changed with respect to the main runs. These runs are presented at Tab.~\ref{tab:subsid}.

   \begin{table}
        \centering
      \caption[]{The 26 subsidiary runs. The smoothing length of
the potential has been set to a smaller value (resp. 20~\% and 40~\% of the disk
thickness) than its usual value (60~\% of the disk thickness). 
}
         \label{tab:subsid}
     $$ 
         \begin{array}{|l|c|c|}
            \hline
\mbox{Run name}&\mbox{corresponding main run}&\mbox{parameter changed}\\
\hline
\mbox{S20R}15_5^i&\mbox{R}15_5^i, i=1\rightarrow 13& \mbox{Smoothing:~} \epsilon=0.2H\\
            \hline
\mbox{S40R}15_5^i&\mbox{R}15_5^i, i=1\rightarrow 13& \mbox{Smoothing:~} \epsilon=0.4H\\
            \hline
         \end{array}
     $$ 
   \end{table}

Subsidiary runs have been performed in  which either the smoothing or the resolution has been changed with
respect to the main runs. The runs with a modified smoothing length are presented at Tab.~\ref{tab:subsid},
whereas the high resolution runs are presented at section~\ref{sec:highres}.

\section{Run results}
\subsection{Generic disk response}
\begin{figure}
   \includegraphics[width=\columnwidth]{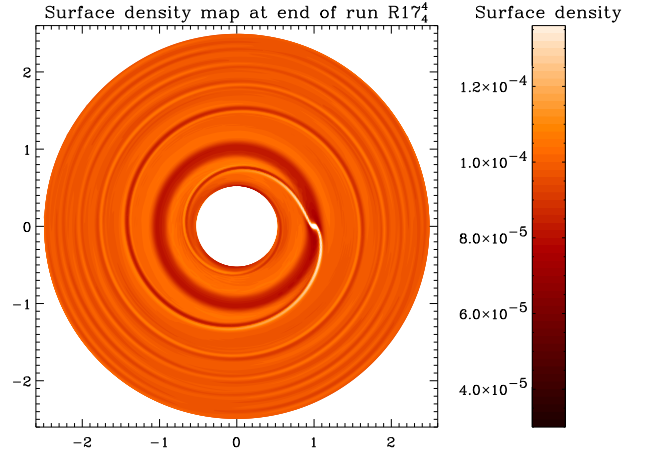}
\caption{\label{fig:sdresp}\new{This plot shows the disk surface density response to the planet
potential, at the end of run~R17$_4^4$. The rotation is anti-clockwise. The one-armed shock triggered by
the planet in the disk is clearly visible, as well as the shallow dip surrounding the orbit, 
the relative depth of which amounts to~$\sim 15$~\% of the disk surface density.}}
\end{figure}
\new{The disk response to the perturbing potential of a planet on a uniform circular orbit is illustrated
at Fig.~\ref{fig:sdresp}. The planet excites waves in the disk at all wavenumbers
(Goldreich \& Tremaine 1979), which corotate 
with the planet, and the superposition
of which leads to a one-armed spiral shock (Goodman \& Rafikov, 2001). This latter is in advance w.r.t. the planet
in the inner disk, which corresponds to the positive inner Lindblad torque, while it is delayed w.r.t the
planet in the outer disk, which corresponds to the negative outer Lindblad torque. The balance between the
one-sided Lindblad torque and the viscous torque, for a steady flow in the planet frame, enables one
to evaluate the depth of the dip which surrounds the orbit (Masset 2001), which, as long as it is shallow, scales
proportionally to the planet mass square and inversely proportionally to the disk viscosity. The dynamics
of the co-orbital region, i.e. the horseshoe region, is better seen in the $(\theta, r-r_p)$ plane, as shown in the
following sections.
}
\subsection{An illustrative example}
\label{sec:example1}
\begin{figure}
   \includegraphics[width=\columnwidth]{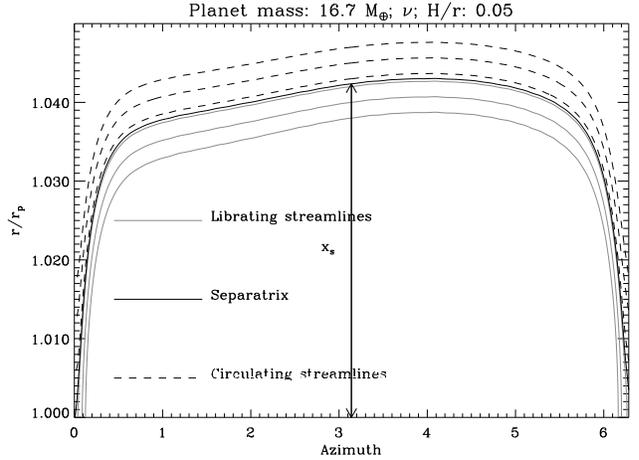}
\caption{\label{fig:sl1bis}Streamlines 
aspect in the co-orbital region for run~R$17_5^1$. 
The thick line represents the outer downstream separatrix.
}
\end{figure}

\begin{figure}
   \includegraphics[width=\columnwidth]{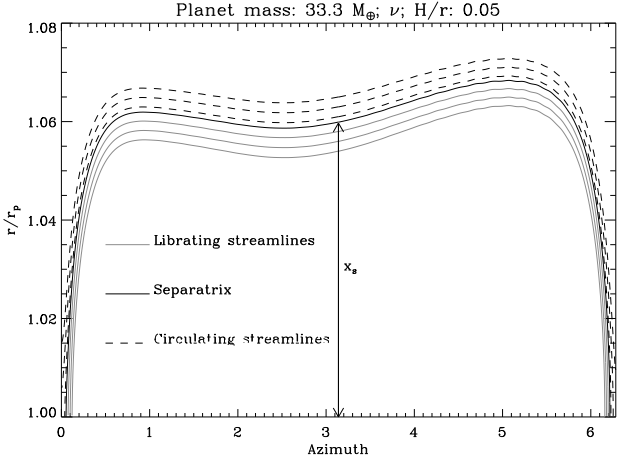}
\caption{\label{fig:sl1}Same as Fig.~\ref{fig:sl1} 
for run~R$33_5^0$. The separatrix lies further from the planet orbit than in run~R$17_5^1$.}
\end{figure}

\begin{figure}
   \includegraphics[width=\columnwidth]{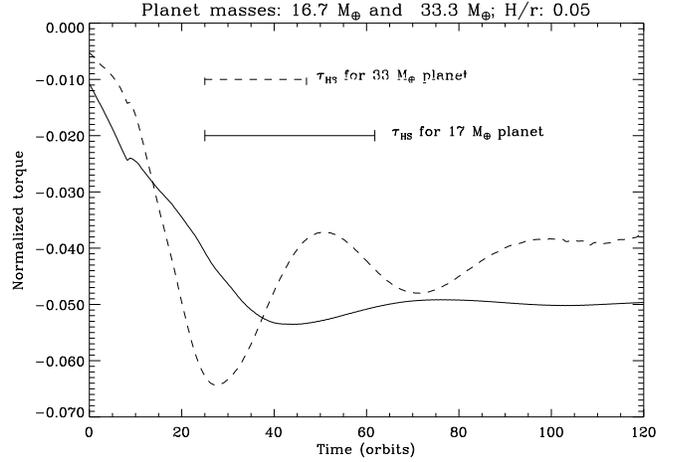}
\caption{\label{fig:tq1}Torque on the planet of run~R$17_5^1$ (solid line) 
and run~R$30_5^0$ (dashed line) 
as a function of time. The outermost horseshoe turnover time is indicated in either case.
The torque is normalized by $\Gamma_0=\pi r_p^4\Omega_p^2\Sigma q^2/h^3$. The glitch around
$t=10$~orbits corresponds to an edge effect of the smoothing window (these results are
smoothed over a temporal window of about 10~orbits).
}
\end{figure}

Let one consider the runs R$17_5^1$ and R$33_5^1$, 
which correspond respectively 
to: $q=5\cdot 10^{-5}$, $h=0.05$ and $\hat\nu=1.93\cdot
10^{-7}$ 
and to: $q=10^{-4}$, $h=0.05$ and $\hat\nu=
1.93 \cdot 10^{-7}$. 
The aspect of the streamlines at time
$t=754\,\Omega_p^{-1}$ in the frame corotating with the planet
in either case
is represented at Fig.~\ref{fig:sl1bis} and~\ref{fig:sl1}.
The streamlines are found integrating the velocity field, which is bilinearly 
interpolated from the values at the interfaces between adjacent zones. As most 
of the motion in the co-orbital region can be accounted for by a linear shear $v_y=2Ax$
where $A$ is the Oort's constant, this bilinear interpolation gives satisfactory results,
and in particular enables one to search for the precise position of the separatrix
between the librating and circulating streamlines. The separatrix position is defined
by its distance $x_s$ to the orbit in opposition with the planet, i.e. at azimuth~$\theta
=\pi$, as indicated in Figs.~\ref{fig:sl1bis} and~\ref{fig:sl1}.
The value of $x_s$ is determined with a dichotomic search. If a guess of~$x_s$ gives
a circulating (resp. librating) streamline, a smaller (resp. higher) value is tried.
The search is iterated until a sufficient precision is achieved. Because the bilinear
interpolation of the velocity field is well adapted to the kinematics of the co-orbital
region, the precision that can be achieved on the separatrix position is much smaller
than the radial zone size, and in that sense such a precision is illusory. However, 
this allows to investigate the behavior of the librating zone width when varying any
parameter of the problem, the grid resolution being fixed.
\new{An estimate of the error on the corotation torque and on the separatrix position due
to finite resolution effect can be found in Appendix~\ref{ap:corot}.}
For the run~R$17_5^1$, the separatrix position is found to be:~$x_s= 0.0424\pm10^{-4}$,
while
for the run~R$33_5^1$, the separatrix position is found to be:~$x_s= 0.0597\pm10^{-4}$,
as can be estimated from Fig.~\ref{fig:sl1bis} and~\ref{fig:sl1}.
Therefore the viscous time-scale across the horseshoe zone half-width is:
\begin{eqnarray}
\tau_{\mathrm{visc}}=\frac{x_s^2}{3\nu}&\simeq& 3.1\cdot 10^3\,\Omega_p^{-1} \mbox{~for R}17_5^1\\
&\simeq&6.2\cdot 10^3\,\Omega_p^{-1}\mbox{~for R}33_5^1
\end{eqnarray}
while the outermost horseshoe turnover time is:
\begin{eqnarray}
\tau_{\mathrm{HS}}=\frac{8\pi}{3\hat x_s}\Omega_p^{-1}&\simeq& 200\,\Omega_p^{-1} \mbox{~for R}17_5^1\\
&\simeq& 140\,\Omega_p^{-1} \mbox{~for R}33_5^1\\
\end{eqnarray}
As the viscous time-scale across the horseshoe region is much higher than the horseshoe
turnover time in either case, one expects the corotation torque to saturate.
The total torque exerted by the disk on the planet as a function of time is represented
at~Fig.~\ref{fig:tq1}.
The total torque tends towards a constant value in time on a time-scale of
the order of the outermost horseshoe turnover time. 
The initial torque value corresponds to the sum of the
differential Lindblad and corotation torque, this later being positive, 
since the surface density is uniform on the co-orbital region. Incidently one can notice that
in both cases
the initial corotation torque value almost cancels out exactly the limit value at large
time,
and that this limit value at large
time is not the same for both planets. 
\begin{figure}
   \includegraphics[width=\columnwidth]{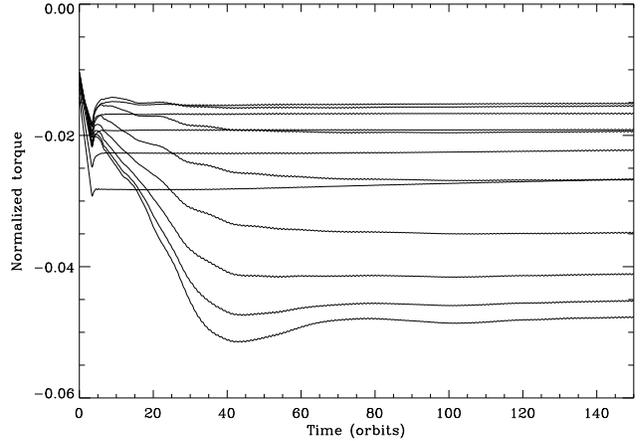}
\caption{\label{fig:tqser}Torque on the planet of runs~R$17_5^i$ 
as a function of time, for $1\leq i\leq 12$. The steady state regime is reached faster as the viscosity increases, and tends
towards a partially saturated state function of the viscosity. The glitch at $t=4$~orbits is an edge effect of the smoothing
window. For this series of runs only the integration time was $150$~orbits in order to check that the total torque value
does not vary significantly after $120$~orbits.
}
\end{figure}
Fig~\ref{fig:tqser} shows the behavior of the total normalized torques as a function of time for the case $q=5\cdot 10^{-5}$,
$h=0.05$, for different viscosities. The bottom curves correspond to low viscosity, and the limit value at large time increases with
viscosity. This limit value reaches a maximum which corresponds roughly to the value at $t=0$, which corresponds to a totally
unsaturated torque, and then decreases beyond the cut-off viscosity, as the topology of the co-orbital region undergoes a change that
will be discussed further (see Masset 2001).

\subsection{Dependence of corotation torque on the viscosity}

\begin{figure}
   \includegraphics[width=\columnwidth]{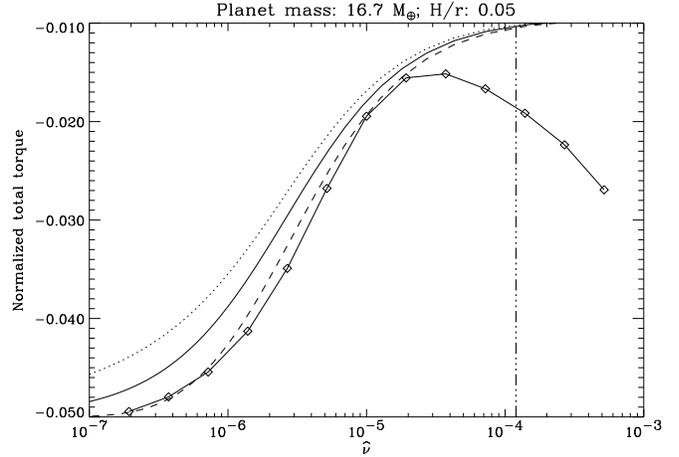}
\caption{\label{fig:tq15}Torque on the planet of runs~R$17_5^i$ (diamonds). The value
of the separatrix distance used for the fits is $x_s=0.0388$.
The vertical three-dot-dashed line
shows the 
cut-off viscosity. Note that the fit maximum value at $-0.01$ is also the torque value at $t=0$ ({\em
cf.} Fig.~\ref{fig:tq1}), and corresponds to a fully unsaturated corotation torque.
}
\end{figure}

Fig.~\ref{fig:tq15} shows the value of the normalized total torque acting
on the protoplanet as a function of the reduced viscosity, evaluated from runs
R$17_5^i$. In each case the value of the torque is averaged on the time interval
$[90,120]$~orbits, in order to get rid of the transitory regime of the beginning,
which corresponds, as we mentioned in section~\ref{sec:example1}, to the time needed for
the corotation torque to (possibly partially) saturate.
This figure also shows the best fits by the functional dependences given by Masset
(2001), which are respectively:
\begin{equation}
\label{eq:gamma1}
\Gamma_C=\Gamma_C^{\mathrm{max}}\cdot \frac 14\left[\frac1{z_s^3}-\frac{g(z_s)}{z_s^4g'(z_s)}\right]
\end{equation}
for the solid line,
\begin{equation}
\label{eq:gamma2}
\Gamma_C=\Gamma_C^{\mathrm{max}}\frac{g(z_s)}{z_sg'(z_s)}
\end{equation}
for the dashed line, and
\begin{equation}
\label{eq:gamma3}
\Gamma_C=\Gamma_C^{\mathrm{max}}\frac{g'(0)}{g'(z_s)}
\end{equation}
for the dotted line, 
where $z_s=\hat x_s(2\pi\hat\nu)^{-1/3}$, where $\Gamma_C^\mathrm{max}=
(9/8)\Sigma_0\Omega_p^2x_s^4$,  and where $g(z)$ is a linear combination of the
Airy functions~Ai and~Bi which cancels out for $z=0$, e.g.:
\begin{equation}
g(z)=\mathrm{Bi}(z)-\sqrt 3\mathrm{Ai}(z)
\end{equation}
The fits presented at~Fig.~\ref{fig:tq15} have the following functional
form:
\begin{equation}
\label{eq:fitfit}
\Gamma(z)=\Gamma_{\mathrm{min}}+\Gamma_C(z)
\end{equation}
where $\Gamma_C$ is one of the three functions given by Eqs.~(\ref{eq:gamma1}),
(\ref{eq:gamma2})
or
(\ref{eq:gamma3})
and therefore has one free parameter, 
the floor level $\Gamma_{\mathrm{min}}$, 
which corresponds to the case of a saturated corotation torque and therefore
to the sum of the differential Lindblad torque and viscosity independent additional term of Eq.~(\ref{eq:add}),
which scales with the one-sided Lindblad torque, and which turns out to be
$\Gamma_{\mathrm{min}}=0.050$.
Although $x_s$ is not a fit free parameter, the strong dependence of $\Gamma_C^\mathrm{max}$ upon $x_s$ and the
relatively high variation  with azimuth of the separatrix distance to the orbit enables one to consider it
as a free parameter within a narrow range (e.g. here in the range $0.038<\hat x_s<0.043$, which can be gathered from
examination of Fig.~\ref{fig:sl1bis}). The value used for the fit of Fig.~\ref{fig:tq15} falls within this range of values, and
it corresponds to the separatrix distance at $\theta\simeq 1.6$~rad. It should be noted that if one discards the lowest viscosity
points for the fit (for which residual numerical viscosity effects may affect the results), then one can get a good fit by 
Eq.~(\ref{eq:gamma1}) as well, with a slightly larger value of $x_s$ which still falls in the range $0.038<\hat x_s<0.043$.
Despite of the uncertainties introduced by the low viscosity points, the fit correctly accounts for a number of properties of the
corotation torque main term (its absolute value, the half saturation for $\hat\nu\sim 0.05\hat x_s^3$, and the fact that the
partial saturation regime
spans about two orders of magnitude of viscosity).

\begin{figure}
   \includegraphics[width=\columnwidth]{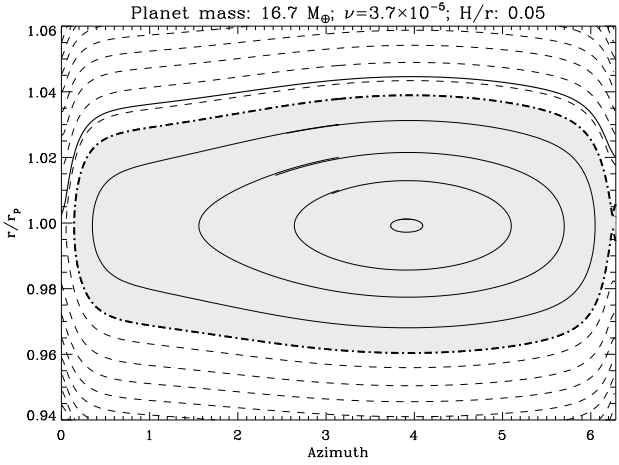}
\caption{\label{fig:highvisc1}
Streamline aspect for the run R$17_5^9$, which corresponds to the peak
value of the total torque of Fig.~\ref{fig:tq15}. The four thin solid lines represent
 librating streamlines. One is almost exactly closed, the other ones not quite, as
a strictly steady state in the planet frame has not been reached yet. The set of 
trapped fluid
elements still have an azimuthal extent which is just smaller than $2\pi$. The dashed
line corresponds to a unique streamline, which originates in the outer disk and eventually
reaches the inner disk. The thick solid line represents the outer downstream separatrix
between circulating and librating streamlines. The shaded area shows the set of librating
fluid elements, the boundary of which (the thick dot dashed line) is also the outer
upstream separatrix  and inner downstream separatrix.
}
\end{figure}

\begin{figure}
   \includegraphics[width=\columnwidth]{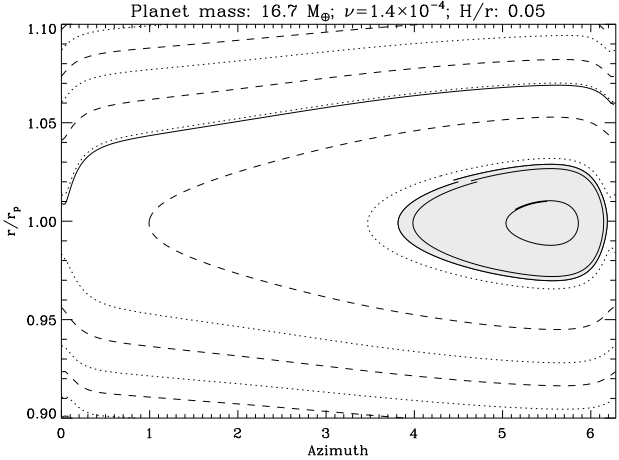}
\caption{\label{fig:highvisc2}
Streamline aspect for the run R$17_5^{13}$. The three thin solid lines represent librating
streamlines. They are approximately closed. The azimuthal extent of the
trapped material (shaded area) is only about $2$~rad, and its boundary roughly coincides with the
outer closed streamline. The dotted and dashed lines represent two open streamlines
originating in the outer disk. The thick solid line represents the outer downstream
separatrix.
}
\end{figure}

The behavior at high viscosities differs markedly from the fit  by equation~(\ref{eq:fitfit}). Indeed a cut-off of the corotation torque
is expected when the viscous drift time of a fluid element across the outer half of the
horseshoe region is shorter than half of its libration time. In that case the region
of fluid elements trapped in the co-orbital region has an azimuthal extent shorter than
the azimuthal extent of the horseshoe region. Figs.~\ref{fig:highvisc1}
and~\ref{fig:highvisc2} represent the streamlines for two cases of high viscosity (resp.
$\hat\nu=3.7\cdot 10^{-5}$ and $\hat\nu=5.2\cdot 10^{-4}$).
The outer downstream separatrix of Fig.~\ref{fig:highvisc1} roughly coincides with that of Fig.~\ref{fig:sl1bis}. The material lying interior
to it is not librating however, as is illustrated by  the dashed streamline. This material undergoes one horseshoe like close encounter with the
planet and goes to the inner disk. The distance between the outer downstream separatrix and the boundary of the trapped region (given
for $r>r_p$ by the outer upstream separatrix) increases with the viscosity and eventually becomes as large as the horseshoe region. In that
case, the material can go from the outer disk to the inner disk without any close encounter with the planet.
Fig.~\ref{fig:highvisc2} shows the example of two open streamlines
(dotted and dashed lines)
which correspond to two fluid elements which reach the inner disk without any 
orbit crossing close
encounter with the planet, a situation that is impossible at low viscosities.
Note that the distance of the outer downstream separatrix to the orbit, measured
for $\theta\simeq 0.75$ (instead of $\theta=\pi$ in order for the viscous drift not to 
affect sizably the measure at these high viscosities), is in the range $[0.035; 0.04]$
in Figs.~\ref{fig:sl1bis}, \ref{fig:highvisc1} and~\ref{fig:highvisc2}.
Therefore
this value is almost constant over three decades of viscosity, which justifies {\it a posteriori}
the choice of a unique value of $\hat x_s$ for the fits of Fig.~\ref{fig:tq15}.

\subsection{Other runs results}
\label{sec:md}
\begin{figure}
   \includegraphics[width=\columnwidth]{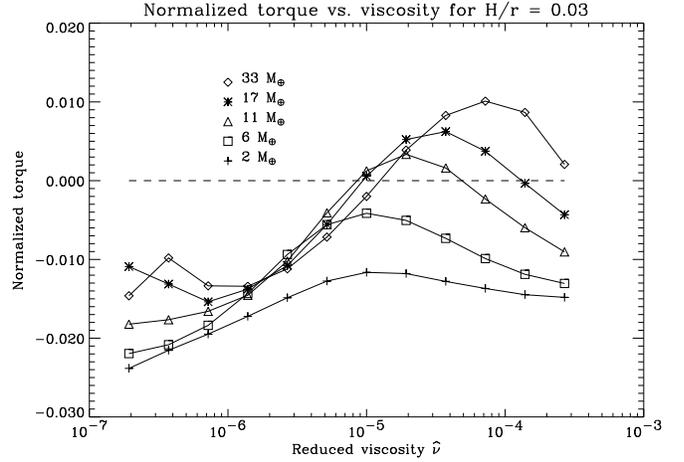}
\caption{\label{fig:h03} Total torque exerted on the planet as a function of viscosity, for a disk aspect ratio
$h=0.03$ and five different values of the planet mass.
}
\end{figure}

\begin{figure}
   \includegraphics[width=\columnwidth]{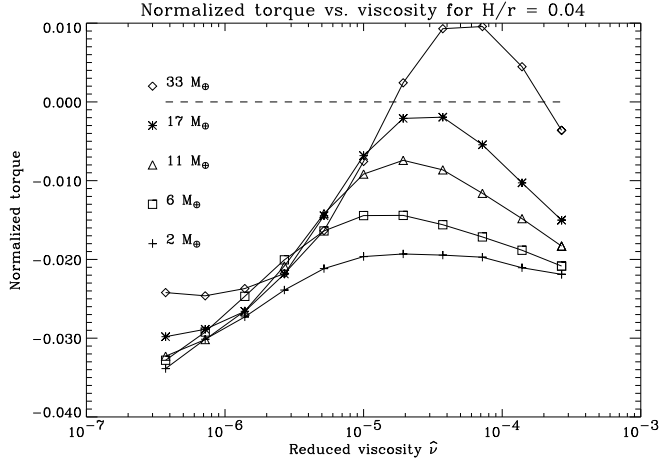}
\caption{\label{fig:h04}Total torque exerted on the planet as a function of viscosity, for a disk aspect ratio
$h=0.04$ and five different values of the planet mass.
}

\end{figure}

\begin{figure}
   \includegraphics[width=\columnwidth]{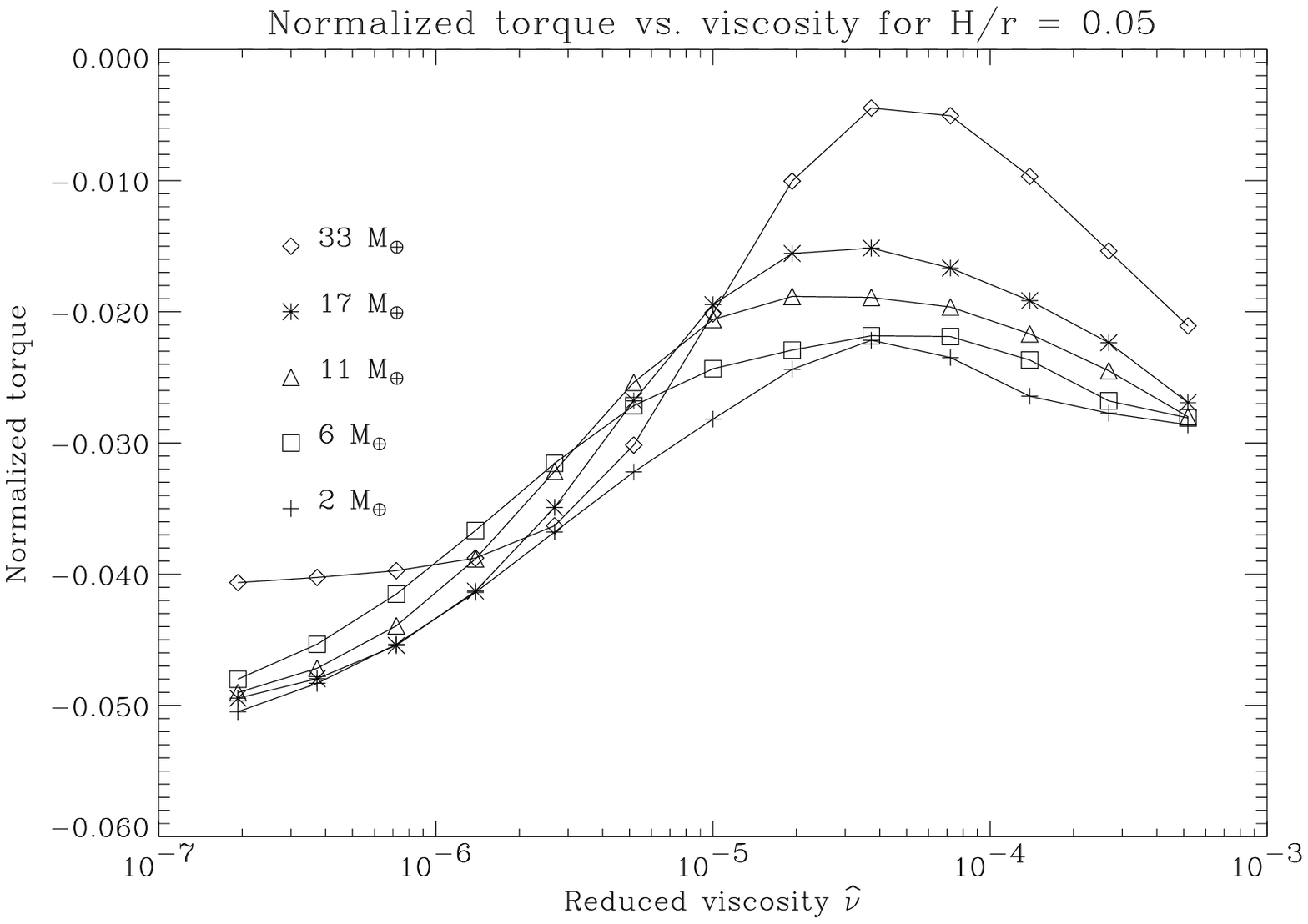}
\caption{\label{fig:h05}Total torque exerted on the planet as a function of viscosity, for a disk aspect ratio
$h=0.05$ and five different values of the planet mass.
}
\end{figure}

\begin{figure}
   \includegraphics[width=\columnwidth]{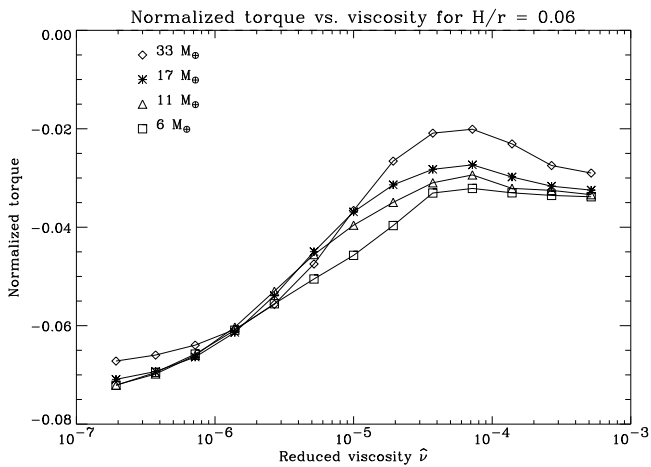}
\caption{\label{fig:h06}Total torque exerted on the planet as a function of viscosity, for a disk aspect ratio
$h=0.06$ and four different values of the planet mass.
}
\end{figure}

Figs.~\ref{fig:h03} to~\ref{fig:h06} show 
the value of the steady state normalized total torque as a function of viscosity
for different planet masses, respectively for disk thickness $h=0.03$,
$h=0.04$,
$h=0.05$, and
$h=0.06$.
It can checked on these figures that the viscosity at which the torque begins to
decrease increases with the planet mass. This is in agreement with the fact that
the cut-off at high viscosity occurs for $\hat\nu\simeq(1/4\pi)\hat x_s^2$, and that
$\hat x_s$ increases with the planet mass.
It can also be checked that the torque saturation occurs at lower viscosities for smaller
masses, as the saturation occurs for $\hat \nu \sim \hat x_s^3$.


\label{sec:dippb}

These results can 
be used to try identify the additional corotation torque term of Eq.~(\ref{eq:add}).
The main problem
is to evaluate the residual torque, i.e. to correct the measured torque from the main corotation torque term.  This latter
involves the separatrix distance to the orbit $x_s$ to the fourth power, and it  is therefore very sensitive
to any error on the estimate of this separatrix distance. Two solutions
can be adopted: the first one consists in considering that the measured torque value at low viscosity is the
sum of the differential Lindblad torque and the additional corotation torque term (in which case there is no need to
evaluate the main corotation torque term) and the second one consists in correcting the torque measured at large viscosity
--although much before the cut-off-- from this main term and attributing the residual to the sum of the differential Lindblad term and the
additional corotation torque term. This latter solution turns out to be easier and more reliable, despite of the difficulty in measuring $x_s$. 
Indeed the disk response at low viscosity may involve a strongly non-linear disk response (the dip around the orbit tends to become
a fully qualified gap), which modifies the differential Lindblad torque in a way that is not trivial to correct. Furthermore, finite numerical viscosity
effects (which are difficult to quantify)
and the long viscous time-scale are two additional factors which may lead to a significant difference between 
the ideal low viscosity steady state
situation and the run results after $120$~orbits. 

\begin{figure}
   \includegraphics[width=\columnwidth]{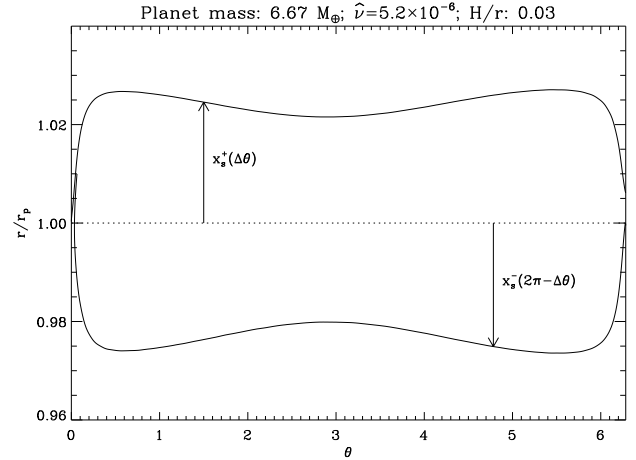}
\caption{\label{fig:exetrang}Separatrix shape for a low mass, low aspect ratio case. The distance of the separatrix to the orbit varies by 
about $23$~\% over the azimuth and therefore its fourth
power varies by a factor $2.3$ with  azimuth. $x_s^+$ and $x_s^-$ stand for the distance to the orbit of the outer and inner separatrices, respectively. They are functions of the azimuth $\theta$.
In this example $\Delta\theta=1.5$~rad.
}
\end{figure}

\begin{figure}
   \includegraphics[width=\columnwidth]{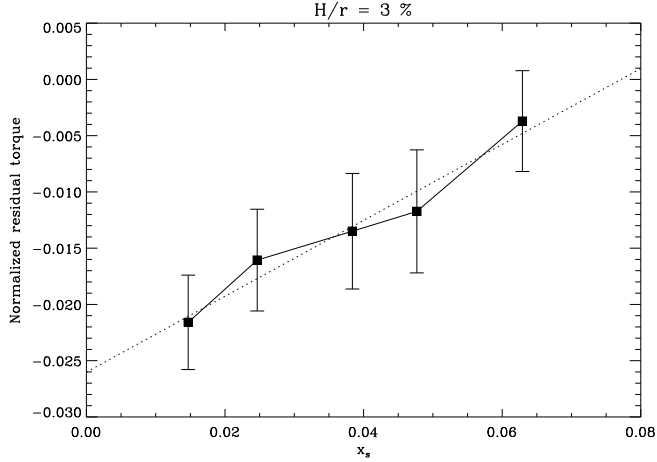}
\caption{\label{fig:cor3}Residual normalized torque as a function of the separatrix distance, for the $3$~\% aspect ratio disk. The dotted line shows the linear regression fit. 
The error bars are obtained by taking
respectively the maximum and minimum value of the average of $x_s^+(\Delta\theta)$ and $x_s^-(2\pi-\Delta\theta)$, when $\Delta$ varies. 
}
\end{figure}

\begin{figure}
   \includegraphics[width=\columnwidth]{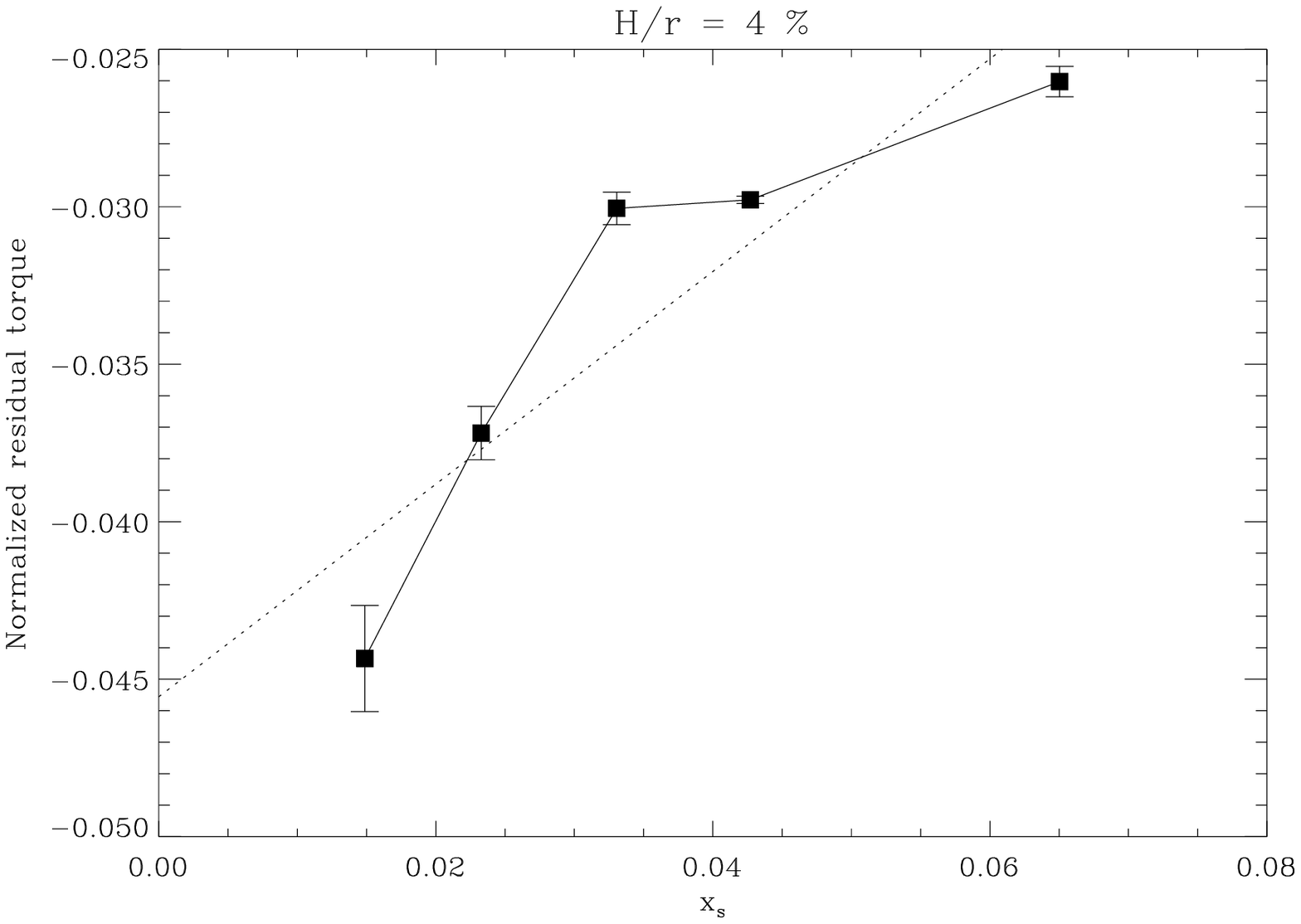}
\caption{\label{fig:cor4}Residual normalized torque as a function of the separatrix distance, for the $4$~\% aspect ratio disk. The dotted line shows the linear regression fit. 
}
\end{figure}

\begin{figure}
   \includegraphics[width=\columnwidth]{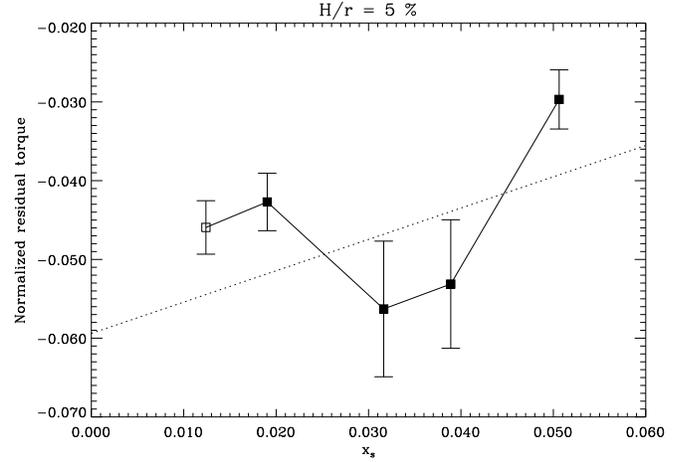}
\caption{\label{fig:cor5}Residual normalized torque as a function of the separatrix distance, for the $5$~\% aspect ratio disk. The dotted line shows the linear regression fit. The white square shows a measurement
for which the separatrix distance to the orbit is smaller than the radial zone width, which means that the librating region is narrower than two zones in radius. It is therefore discarded for the linear regression fit.
}
\end{figure}

\begin{figure}
   \includegraphics[width=\columnwidth]{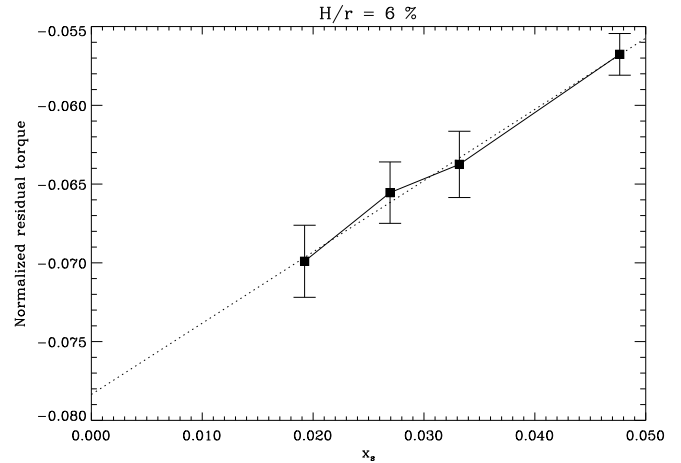}
\caption{\label{fig:cor6}Residual normalized torque as a function of the separatrix distance, for the $6$~\% aspect ratio disk. The dotted line shows the linear regression fit. 
}
\end{figure}

Evaluating the corotation torque main term involves a precise measurement of the value of $x_s$. It can be seen in Fig.~\ref{fig:exetrang}
how the distance
between the separatrix and the orbit varies with azimuth. 
The value for $x_s$ is chosen to be the arithmetical average of the
outer downstream separatrix position at a given angle $\Delta\theta$ from outer conjunction and the inner downstream separatrix position 
at the corresponding angle $2\pi-\Delta\theta$ from inner conjunction:
\begin{equation}
x_s=\frac 12[x_s^+(\Delta\theta)+x_s^-(2\pi-\Delta\theta)]
\end{equation}
The value of $\Delta\theta$ is then varied between~$1$ and~$\pi$~rad, which leads to a dispersion in the corotation torque main
term evaluation, and this leads therefore to a dispersion of the residual torque estimate. The corresponding residual torque estimates
are shown for the four different disk thicknesses in Figs.~\ref{fig:cor3} to~\ref{fig:cor6}.

The results of the linear regression fits performed on these figures are shown in Tab.~\ref{tab:addterm}.
A number of comments can be made from these figures and from the table:
\begin{enumerate}
\item The correction is performed at the peak value of the total term, while the separatrix position is searched for at a slightly
lower viscosity. The correction takes into account a possibly
partial saturation of the corotation torque at its peak value.
\item The function which is used to correct the main corotation term from partial saturation is the function given by 
Eq.~(\ref{eq:gamma1}). Inspection of Fig.~\ref{fig:tq15} shows that using either
Eq.~(\ref{eq:gamma2})
or Eq.~(\ref{eq:gamma3}) would not change significantly the results, given the uncertainty on $x_s$, and given the fact that at its
peak value the corotation torque is almost fully unsaturated.
\item The value at $x_s=0$ of the linear regression fit corresponds roughly to the limit value that can be deduced by eye from the limit value at $\hat \nu=0$ in Fig.~\ref{fig:h03} and~\ref{fig:h04}.
In the case of $h=0.05$, the uncertainties are considerable, and in the $h=0.06$ case, no curve reaches the fully saturated regime at 
low $\hat\nu$.
\item The mean slope $\overline{\cal G}$ distribution is compatible with the expectation. It should be mentioned however that the function ${\cal G}(x_s)$ is not expected to be a constant,
and that its value, which is such that: $1\leq{\cal G}(x_s)\leq7/3$, depends on the location $x_s$ at which the separatrix samples the axisymmetric profiles perturbations.
\item The variation with mass of the peak value of the total normalized torque for a given aspect ratio can 
be accounted for by the slope of the
fit, i.e. this variation is roughly equal to $\overline{\cal G}(\hat x_s^\mathrm{max}-\hat x_s^\mathrm{min})\Gamma_\mathrm{LR}$. This
therefore accounts for the increasing peak value with the planet mass, whereas one would expect a constant maximum in the linear regime
($x_s^4\propto q^2$) and a decreasing peak value beyond ($x_s^4 \propto q^{4/3}$). 
\item Not all the behavior observed is entirely due to the additional corotation torque term. Indeed
there is no reason to assume that the normalized
differential Lindblad torque is a constant for a given aspect ratio. Although this is certainly true in the linear regime, it is no
longer true
for higher masses. Miyoshi et al. 1999 note that in an infinitesimally thin disk, in the non-linear regime (the threshold
of which they find at $R_H = H/2$, where $R_H$ is the planet Hill radius), the one-sided Lindblad torque is smaller than its linear
estimate. They mention that part of the cut-off comes from material inside the horseshoe region, which exerts an opposite torque on the
perturber, and weakens the linear estimate of the Lindblad torque. As the material in the horseshoe region is librating in the planet frame,
this material in average corotates with the planet, and therefore participates in the corotation torque and no in the Lindblad torque. If one
defines the Lindblad torque as the torque exerted by the circulating material on the planet, which is, in the steady state regime, the only
definition which ensures that the total torque exerted on the planet is the sum of the corotation and Lindblad torques (as the co-orbital
dynamics partitions the disk into librating and circulating fluid elements) then it is not clear whether the behavior observed by Miyoshi et al.
is due to a Lindblad torque non-linear cut-off or to an avatar of the additional corotation torque term. If one performs however a local scattering
calculation (Lin \& Papaloizou 1979, Papaloizou \& Lin 1984) over the circulating fluid elements, then one expects a $q^{-1}$ cut-off
of the one-sided Lindblad torque when the separatrix lies further than $(2/3)H$ from the orbit, i.e. when the horseshoe region invades
what would be otherwise the place of disk-perturber angular momentum exchange through Lindblad torques. It should be noted that
on the present data set, the conditions $x_s > (2/3)H$ and $R_H >(1/2)H$ are almost strictly equivalent. In the $h=3$\% case
(Fig.~\ref{fig:cor3}), all points but the bottom one correspond to situation beyond the non-linear threshold, while in the $h=6$\% case
(Fig.~\ref{fig:cor6}), all points correspond to a situation where the one-sided Lindblad torque is still in the linear regime (the Hill radius
of the most massive planet is precisely $0.03r_p$). Therefore the behavior observed for the thinner disks is likely to be contaminated
by a differential Lindblad torque non-linear cut-off (which also conspires to lift the normalized curves as the planet mass increases), 
whereas the
behavior observed for the $h=0.06$ disk is exempt of this contamination and should be due entirely to the additional corotation torque
term of Eq.~(\ref{eq:add}).
\end{enumerate}

   \begin{table}
        \centering
      \caption{Linear fit regression values for Figs.~\ref{fig:cor3} to~\ref{fig:cor6}. The slope $\overline{\cal G}$ is defined by
$\Gamma(\hat x_s)=\Gamma_{\hat x_s\rightarrow 0} + \overline{\cal G}\hat x_s\Gamma_\mathrm{LR}$, where
the normalized value of $\Gamma_LR$ for all the disks is $0.21$.
}
         \label{tab:addterm}
     $$ 
         \begin{array}{|c|c|c|c|}
            \hline
        {h (\%)} & \mbox{slope }\overline{\cal G}{\phantom{\int^W}} & \Gamma_{\hat x_s\rightarrow 0} & \mbox{correlation}\\
\hline
        3       & 1.61  &       -0.026  &       0.978 \\
        4       & 1.61  &       -0.046  &       0.90  \\
        5       & 1.89  &       -0.060  &       0.44 \\
        6       & 2.15  &       -0.077  &       0.997 \\
            \hline
         \end{array}
     $$ 
   \end{table}

Despite of the difficulty of these measures, 
these results confirm the existence of a term in the total torque which scales roughly as  $\hat x_s\Gamma_\mathrm{LR}$, 
in agreement with the expectation that ${\cal G}(x_s)=O(1)$, and ${\cal G}(x_s) < 7/3$.

\section{Smoothing issues and 3D expectations}
\label{sec:smoothing}
The above simulations are performed in 2D, and the disk finite thickness is taken into
account through the use of a smoothing coefficient~$\epsilon$ in the potential, where
$\epsilon$ is a sizable fraction of the disk vertical scale length~$H=r\frac{c_s}{v_K}$.
One can wonder how accurate this description is, and what fraction $\eta=\epsilon/H$ of the disk
thickness should be used in order to get as close as possible a result to the
3D expectations. Moreover, there is no reason that an adequate value of $\epsilon$ for the
Lindblad torque (in the sense that this value of $\epsilon$ will provide a value for the
one-sided and/or differential Lindblad torque in agreement with results obtained with
three dimensional calculations, see e.g. Miyoshi et al. 1999) will also be an adequate value
for the corotation torque. 
The following discussion is therefore divided in two parts:
the effect of smoothing on the Lindblad torque is first investigated,  then secondly on the corotation torque, and lastly it is discussed
whether Lindblad and corotation torques can be described appropriately by the
same smoothing coefficient.
\subsection{Effect of smoothing on the Lindblad torques}
\begin{figure}
   \includegraphics[width=\columnwidth]{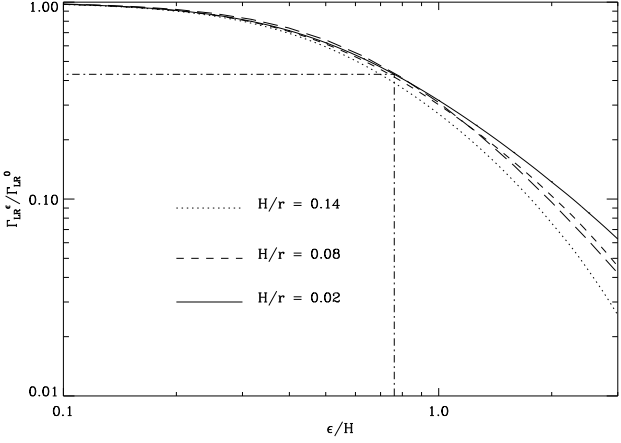}
\caption{\label{fig:onesidedLTsm}One-sided Lindblad torque cut-off as a function of the relative smoothing
$\epsilon/H$. The long dashed line represents the approximate asymptotic solution for $h\rightarrow 0$
deduced from  Eq.~(\ref{eq:asym}).
}
\end{figure}

Fig.~\ref{fig:onesidedLTsm} shows the ratio of the one-sided Lindblad torque for
a smoothed potential and the one-sided Lindblad torque for an unsmoothed potential, as a
function of the ratio of the smoothing length to the disk thickness. The one-sided Lindblad torque is 
 evaluated by the arithmetical average of the outer and inner Lindblad
torque expressions, which are:
\begin{equation}
\Gamma_\mathrm{OLR}=\sum_{m=0}^{+\infty}\Gamma_+^m
\end{equation}
and:
\begin{equation}
\Gamma_\mathrm{ILR}=\sum_{m=0}^{+\infty}\Gamma_-^m
\end{equation}
where:
\begin{equation}
\label{eq:indim}
\Gamma_\varepsilon^m=-\varepsilon\pi m\frac{[\psi_m^\varepsilon(h,\epsilon)]^2h^
3}{(rdD/dr)_{r_m^\varepsilon}}
\end{equation}
in which $\varepsilon=\pm 1$ (for Outer/Inner Lindblad resonance respectively), and
where:
\begin{equation}
\label{eq:tqq}
\psi_m^\varepsilon(h,\epsilon)=\frac{\left(r\frac{d}{dr}G_m^\epsilon(r)\right)_{r_m^\varepsilon}+
2m^2\frac{\Omega_m^\varepsilon-\Omega_p\sqrt{1-h^2}}{\Omega_m^\varepsilon}G_m^\epsilon(r_m^\varepsilon)}{(1
+4m^2h^2)^{1/2}}
\end{equation}
in which $G_m^\epsilon(r)$ is a generalized Laplace coefficient (i.e. a Laplace coefficient for which a smoothing
is introduced):
\begin{equation}
\label{eq:laplsm}
G_m^\epsilon(r)=\frac2\pi\int_0^\pi\frac{\cos(m\theta)}{(1-2r\cos\theta+r^2+\epsilon^2)^{1/2}}
d\theta
\end{equation}
and where:
\begin{equation}
\label{eq:omega}
\Omega_m^\varepsilon=\Omega_p[(1-h^2)^{1/2}+\varepsilon m^{-1}(1+m^2h^2)^{1/2}]^{-1}
\end{equation}
is the Keplerian frequency at the effective location of the Lindblad resonances,
and:
\begin{equation}
r_m^\varepsilon=\left(\frac{\Omega_m^\varepsilon}{\Omega_p}\right)^{-2/3}
\end{equation}
and we use:
\begin{equation}
\label{eq:lastw}
\left(r\frac{dD}{dr}\right)_{r_m^\varepsilon}=-3m\varepsilon\Omega_m^\varepsilon
\Omega_p(1+m^2h^2)^{1/2}
\end{equation}

This calculation is identical to the calculation in Ward (1997) except for the introduction
of the smoothing of the perturbing potential.

\begin{figure}
   \includegraphics[width=\columnwidth]{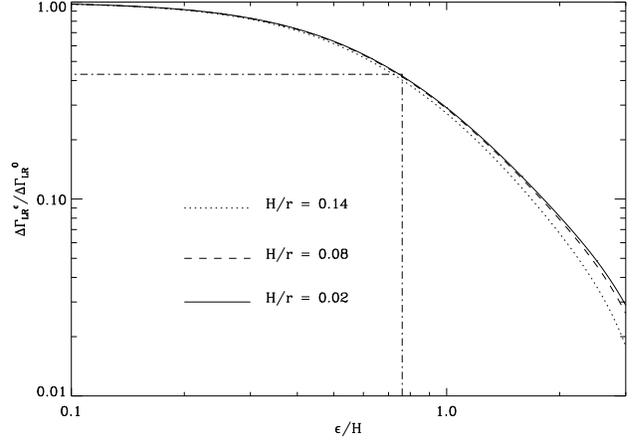}
\caption{\label{fig:diffLTsm}Differential Lindblad torque cut-off as a function of the relative smoothing
$\epsilon/H$.
}
\end{figure}

The cut-off  function is shown for three different aspect ratios, and found to be only weakly dependent
on~$h$. At low aspect ratio, the functional dependence of the torque cut-off tends towards
a fixed function, the graph of which should roughly coincide with the solid curve of 
Fig.~\ref{fig:onesidedLTsm}.  An approximation of the  limit cut-off function when $h\rightarrow 0$ can be
obtained by developing Eq.~(\ref{eq:laplsm})
 to first order in $mh$, and 
approximating the $G_m^\epsilon$ coefficients
with  a standard technique as a function of the Bessel K$_0$ and K$_1$ functions. The
summation over $m$ can then be approximated as an integral. Using the variables $\xi=mh$ and $\eta=
\epsilon/H$, one can finally write the following approximate formula for the one-sided Lindblad
torque as a function of the relative smoothing~$\eta$ only:
\begin{equation}
\label{eq:asym}
\gamma_\mathrm{LR}(\eta)=K\int_0^\infty\xi^2\frac{(1+\xi^2)^{1/2}}{1+4\xi^2}
F\left[\frac 49(1+\xi^2)+\xi^2\eta^2\right]d\xi
\end{equation}
where $K$ is a constant and:
\begin{equation}
F(x)=\frac{1}{3\sqrt{x}}\mathrm{K}_1(\sqrt{x})+\mathrm{K}_0(\sqrt{x})
\end{equation}
where $\mathrm{K}_0$
and $\mathrm{K}_1$ are the modified Bessel functions. The long dashed line of Fig.~\ref{fig:onesidedLTsm} is 
the graph of the cut-off $\gamma_\mathrm{LR}(\eta)/\gamma_\mathrm{LR}(0)$.
Fig.~\ref{fig:diffLTsm} shows the behavior of the differential Lindblad torque cut-off for the same set of disk thicknesses.
This cut-off exhibits quite similar a behavior to the one-sided Lindblad torque cut-off.
On Figs.~\ref{fig:onesidedLTsm} and~\ref{fig:diffLTsm} the dot-dashed line shows the value of the
smoothing for which the torque cut-off amounts to $0.43$, which is the value quoted by Miyoshi et al.~1999
for the ratio of the Lindblad torque in a vertically resolved disk and in an infinitesimally thin disk,
in the linear regime. The corresponding relative smoothing is $\eta=0.76$, both on the one-sided Lindblad
torque and on the differential Lindblad torque. Therefore a potential smoothing length of $76$~\% of the disk
thickness should be used in 2D simulations in order to give correct results for the (differential) Lindblad
torque. The fact that this value is independent of the disk aspect ratio and planet mass (at least in the
linear regime) comes from the fact that most of the torque between the disk and the planet comes from a
zone which is at a distance $\pm H$ from the orbit (the Lindblad resonances pile up for $m\rightarrow\infty$
at a distance $\pm \frac 23H$ from the orbit). Such an argument cannot be used for the co-orbital corotation 
torque which needs a different treatment.

\subsection{Effect of smoothing on the corotation torque}
\begin{figure}
   \includegraphics[width=\columnwidth]{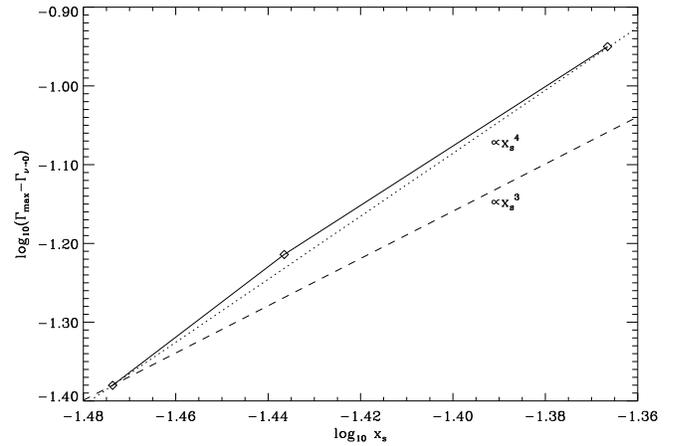}
\caption{\label{fig:xs4}Corotation torque main term estimate as a function of the separatrix position. The dotted line correspond
to a fourth power of $x_s$, and the dashed one to a third power of $x_s$.}
\end{figure}

The corotation torque comes from the exchange of angular momentum between the planet and nearby,
orbit crossing fluid elements (which can be either fluid elements librating on closed, horseshoe like
streamlines, or fluid elements participating in the global viscous accretion of the disk, and which
pass by the planet  as they go from the outer disk to the inner disk). As a fluid element participating
in the corotation torque,
with impact parameter $x=r-r_p$, will be located at a distance $-x$ from the orbit after a back-scattering
by the planet, the specific angular momentum that it gives to the planet is $4Bxr_p$, and therefore this value
does {\em not} depend on smoothing. The total value of the corotation torque does depend on it however,
since the radial width of the libration region depends on the smoothing: the smaller the smoothing,
the further the separatrices lie from the orbit, and the larger the corotation torque.
\begin{figure}
   \includegraphics[width=\columnwidth]{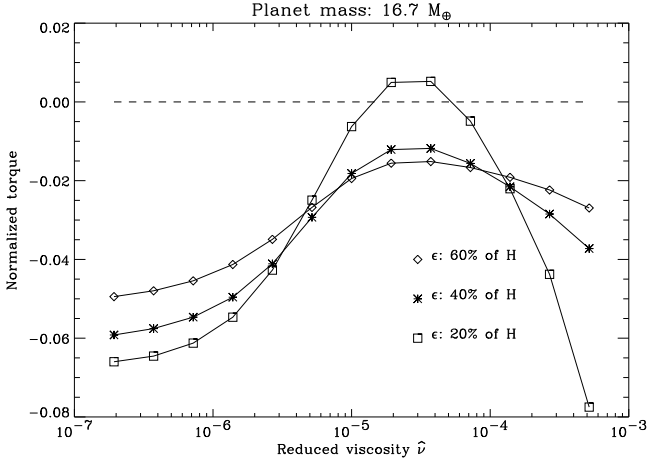}
\caption{\label{fig:corsm}Total torque acting on the planet, as a function of the reduced viscosity,
for runs
$\mbox{S20R}17_5^i$,
$\mbox{S40R}17_5^i$
and R$17_5^i$.}
\end{figure}
This behavior can be checked at Fig.~\ref{fig:corsm}. The runs for a planet of mass $m_p=16.7$~$M_\oplus$
have been performed with three different smoothing values.  At low viscosity (i.e.
for a saturated corotation torque) the total torque is negative and has a larger absolute value for a smaller
smoothing. This corresponds to expectations since the torque in this region (whether it
includes the Lindblad torque coupling term of the corotation torque or not) scales as the 
(one-sided or differential) Lindblad torque. On the other hand, the difference between the maximum value of the
torque with the minimum value at low viscosity should roughly correspond to the maximum value 
of the
corotation torque main term, which scales as $x_s^4$. At Fig.~\ref{fig:xs4} we plot this difference as a function of
the separatrix distance to the orbit $x_s$.
Despite the small number of measurements, the results are in good agreement with a $x_s^4$ dependency
of the corotation torque. This also confirms the argument exposed above 
that the corotation torque depends on the smoothing only through
the value of $x_s$. 

It is worth emphasizing the dramatic dependence of the corotation torque on smoothing. 
In Fig.~\ref{fig:corsm}, for a smoothing length that is $60$~\% of the disk thickness,
the torque is always negative, although it can reach a value one order of magnitude smaller than the
differential Lindblad torque linear estimate for an unsmoothed potential. On the other hand, a smoothing
length that is $20$~\% of the disk thickness leads to a torque reversal over a significant range
of viscosity ($1.5\cdot10^{-5} < \hat\nu < 6\cdot 10^{-5}$, which corresponds to $6\cdot 10^{-3}<\alpha<0.024$).

An idea of the co-orbital dynamics in a three dimensional situation can be obtained simply if one assumes
that the motion is purely horizontal. In that case the motion in the slice of altitude $z$ and infinitely
small thickness $dz$ is equivalent to a 2D situation, in which the potential smoothing length is $z$. Furthermore,
as for $|z/r_p| \ll 1$ the specific angular momentum radial gradient is still 
$2Br=\Omega_Kr/2$, the torque contribution
of the slice is the same as the torque exerted by an infinitesimally thin disk (with a potential smoothed over
a length $z$) of surface density $\rho(z)dz$. Let  $X_s(\epsilon, q, h)$ be the separatrix distance to the orbit
in an infinitesimally thin disk in the equatorial plane, in which the potential is smoothed over a length $\epsilon$, the sound speed
is $c_s=v_Kh$, and the planet to primary mass ratio is $q$. The fully unsaturated corotation torque main term 
in a thick disk can therefore be written as:
\begin{equation}
\label{eq:3d1}
\Gamma_c=\int_{-\infty}^{+\infty}\frac 98\Omega_p^2\rho(z)X_s^4\left[z,q,\frac{c_s(z)}{v_K}\right]dz
\end{equation}
As this analysis is restricted to the case of isothermal Keplerian disks, Eq.~(\ref{eq:3d1}) can be recast as:
\begin{equation}
\label{eq:3d2}
\Gamma_c=\int_{-\infty}^{+\infty}\frac 98\Omega_p^2\rho_0e^{-\frac 12(z/H)^2}X_s^4(z,q,h)dz
\end{equation}
where $\rho_0$ is the disk density in the equatorial plane. Eq.~(\ref{eq:3d1}) can also be written as:
\begin{equation}
\label{eq:3d3}
\Gamma_c=\frac 98\Omega_p^2\Sigma_0\overline{x}_s^4
\end{equation}
in which $\overline x_s^4$ is defined as:
\begin{equation}
\overline x_s^4=\int_{\infty}^{+\infty}\frac{\rho(z)X_s^4(z,q,h)}{\Sigma_0}dz
\end{equation}
which can be recast as:
\begin{equation}
\label{eq:3d4}
\overline x_s^4=\frac{1}{\sqrt{2\pi}h}\int_{\infty}^{+\infty}e^{-\frac 12(z/H)^2}X_s^4(z,q,h)dz
\end{equation}
Therefore, if one wants a 2D simulation to give the correct amplitude for the fully unsaturated corotation
torque main term, one needs to use the smoothing length which endows the co-orbital motion with a separatrix
position $\overline x_s$, where $\overline x_s$ is given by Eq.~(\ref{eq:3d4}). The integrand of the
right hand side contains the function $X_s$, which has to be tabulated a priori with a set of runs with different
smoothing lengths, for a given planet mass and disk aspect ratio. The same tabulation is used to infer
the correct value for $\epsilon$ once $\overline x_s$ is known.
\begin{figure}
   \includegraphics[width=\columnwidth]{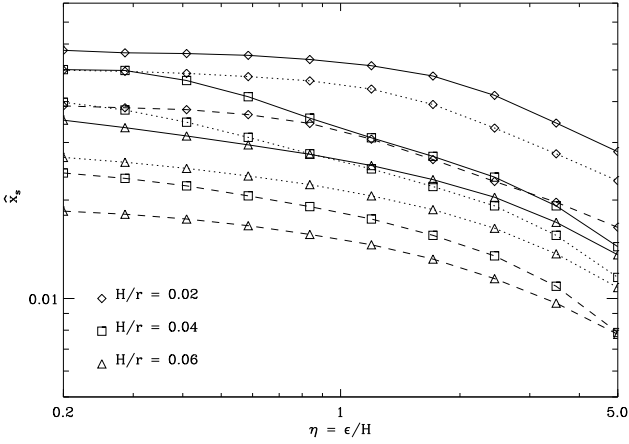}
\caption{\label{fig:xstab}Separatrix distance to the orbit as a function of the relative smoothing 
$\eta=\epsilon/H$, for different aspect ratios and planet masses 
($q=1.67\cdot 10^{-5}$: dashed lines,$q=3.33\cdot 10^{-5}$: dotted lines, and $q=5\cdot 10^{-5}$:
solid lines).  As expected, $x_s$ decreases as the smoothing increases, decreases as the aspect ratio
increases (everything else being fixed), and increases as the mass increases (everything else being fixed).}
\end{figure}
Fig.~\ref{fig:xstab} shows such tabulations for different planet masses and disk aspect ratios. They are
measured with a dichotomic search of the separatrix position on the results of low viscosity runs ($\hat
\nu=10^{-6}$), over $150$~orbits.
\begin{figure}
   \includegraphics[width=\columnwidth]{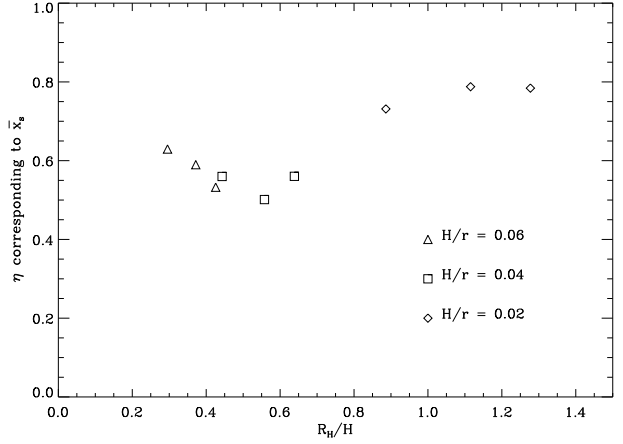}
\caption{\label{fig:etacor}Corotation torque adapted smoothing as a function of $R_H/H$, inferred from the tabulations of Fig.~\ref{fig:xstab}.}
\end{figure}
Fig.~\ref{fig:etacor} shows the results of this integration for the nine situations of Fig.~\ref{fig:xstab}. It can be seen that on the contrary to
the case of the Lindblad torque, the value of the correct smoothing depends on $R_H/H$ even when this value is below $1/2$. Furthermore,
in this case, which corresponds to the linear regime for the Lindblad torque, the correct smoothing length is clearly smaller (50--60~\% of the
disk thickness) than the correct smoothing length for the Lindblad torque (76~\% of the disk thickness). The choice of any value between
$60$ and $76$\% of the disk thickness will therefore underestimate the corotation torque and overestimate the Lindblad torque,
and therefore in any case will underestimate the total torque, since the differential Lindblad torque is negative
and the corotation torque is positive.
Furthermore, as the corotation torque is a very sensitive function of the smoothing, there is little hope of finding
a correct smoothing prescription which predicts a correct total torque value in a 2D simulation, especially in the cases of interest where the
corotation torque and differential Lindblad torque almost cancel out. One can however make conservative assumptions to infer the direction
of migration of a protoplanet by choosing a lower or upper limit for the smoothing coefficient. This is the object of the next section which
presents four series of selected high resolution runs.

\section{Higher resolution runs}
\label{sec:highres}
As the results of the $h=0.03$ and $h=0.04$ runs suggested that it might be possible to reverse migration over a certain viscosity range for the
most massive non-accreting planets ($10\;M_\oplus$ and $17\;M_\oplus$), the corresponding runs have been repeated with  slightly
different values of the smoothing following the discussion of section~\ref{sec:smoothing}, and with a higher resolution in order to rule out
possible finite resolution effects, which could particularly affect the differential Lindblad torque
\new{in the thinnest disks}. Namely, the setups detailed at
Tab.~\ref{tab:hr} have been run.

   \begin{table}
        \centering
      \caption[]{\label{tab:hr}
The series of runs listed below all have $N_r=286$ and $N_\theta=900$, the same $R_\mathrm{max}$ as the previous runs while 
$R_\mathrm{min}$
has been slightly modified and set to $0.509280742$ in order for the planet to be at the center of a zone. The set of reduced viscosities
is in each case: $\hat\nu=5\cdot 10^{-6}\times 20^{i/4}$, $i\in[0,4]$. The last column shows the adopted
ratio $\eta$ of the smoothing length to the disk 
thickness.
}
     $$ 
         \begin{array}{|l|c|c|c|c|}
            \hline
\mbox{Run name}&\mbox{Aspect ratio}&\mbox{Planet mass}&R_H/H&\mbox{$\eta$}\\
\hline
\mbox{HR}11_3^i&        0.03    & 3.33\cdot 10^{-5}&    0.74 &   0.76\\
            \hline
\mbox{HR}11_4^i&        0.04    & 3.33\cdot 10^{-5}&    0.56 &  0.55\\
            \hline
\mbox{HR}17_3^i&        0.03    & 5\cdot 10^{-5}&               0.85 &  0.76\\
            \hline
\mbox{HR}17_4^i&        0.04    & 5\cdot 10^{-5}&               0.64 &  0.55\\
            \hline
         \end{array}
     $$ 
   \end{table}

\begin{figure}
   \includegraphics[width=\columnwidth]{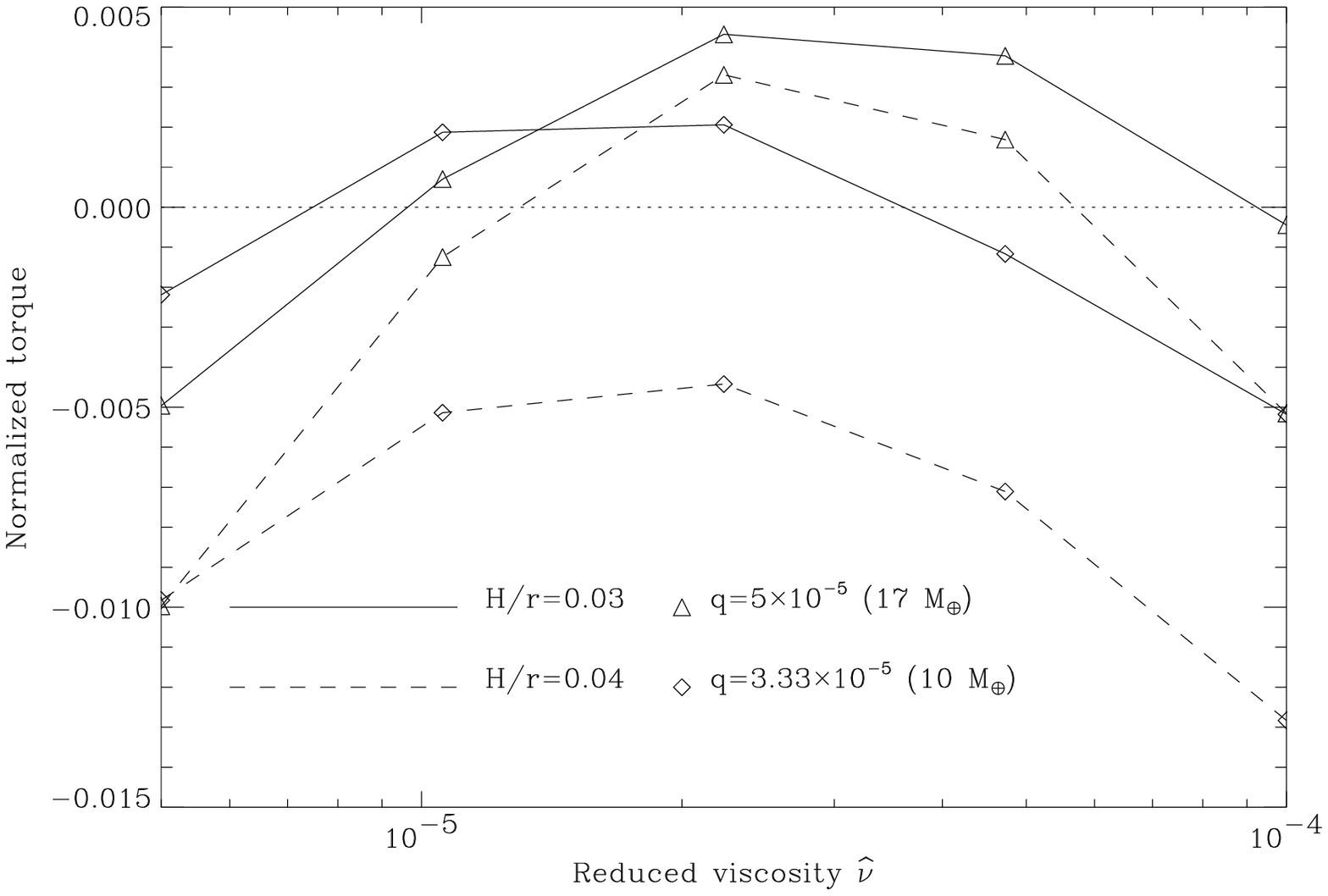}
\caption{\label{fig:hrcurve}}Normalized total torque acting as a function of viscosity for runs HR11$_3^i$,
HR11$_4^i$, HR17$_3^i$ and HR17$_4^i$.
\end{figure}

Each setup has been run for the five values of the viscosity indicated at Tab.~\ref{tab:hr}, ranging from $\hat\nu=5\cdot 10^{-6}$
to $\hat\nu=10^{-4}$, over $60$~orbits only. This amount of time may seem a bit short, but a look at Fig.~\ref{fig:tqser} shows that for
these relatively high values of the viscosity, the limit torque \new{value} at large time is reached \new{on a shorter
time-scale} than the outermost
horseshoe turnover time. This turns out to be also the case with these high resolution runs. 
The value for the smoothing length has been chosen according to the value of the ratio of the Hill sphere radius to the disk
thickness $R_H/H$ and to the results of Fig.~\ref{fig:etacor}. For the largest ratios, a smoothing length of $76$\% of the disk thickness
has been chosen. This value is close to the correct one as far as the corotation torque is concerned, and it is assumed that
it is also the correct one for the Lindblad torque, although such values of $R_H/H$ do not correspond to the linear regime for
which the \new{canonical smoothing} value was obtained. 
One the other hand, for the two smallest $R_H/H$ ratios, a value $\eta=0.55$ was adopted. This
value should lead to correct results as far as the corotation torque is concerned, and should lead to an overestimation of the 
(negative) differential Lindblad torque, therefore they are conservative in the sense that  they give a lower limit on the total torque peak value.
Fig.~\ref{fig:hrcurve} shows the normalized
total torque at large time as a function of viscosity for the four series of runs. 
For the runs 
HR$17_3^i$,
HR$11_3^i$ and
HR$17_4^i$,
the total torque reaches a positive peak value, corresponding to an outwards migration.  The limit $\alpha$ values of the interval
over which the migration is outwards are respectively: $0.01<\alpha<0.1$, $7\cdot 10^{-3} < \alpha < 4\cdot 10^{-2}$, and
$8\cdot 10^{-3}<\alpha<0.037$. In this last case it should be remembered that the choice of the smoothing coefficient was a conservative
one, and therefore it is likely that the actual viscosity interval over which migration reverses is larger. The case of the run series
HR$11_4^i$ shows that at its peak value, around $\alpha\sim 0.01$, the torque value is $22$~times smaller than the linear differential
Lindblad torque estimate for an unsmoothed potential (and this again is a conservative estimate).

It is noteworthy that the results of runs
HR$17_3^i$ and HR$11_3^i$, despite of the small aspect ratio, do not differ much from the results obtained at lower resolution.
\new{This is an indication that the corotation and differential
Lindblad torques were already satisfactorily described at that resolution.}

These results indicate that the migration of Neptune-sized planets (as well as slightly smaller objects) is reversed in thin ($h\leq0.04$),
viscous disks (with viscosities typically between $0.01<\alpha <0.04$, although it should be remembered that the exact extent of this
range depends on the planet mass and disk aspect ratio).

\section{Discussion}
\begin{figure}
   \includegraphics[width=\columnwidth]{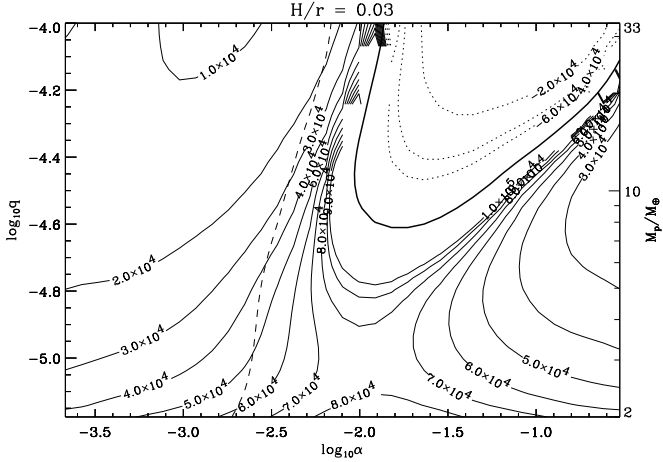}
\caption{\label{fig:taumig3}Migration time in years at~1~AU for a $h=0.03$ disk, as a function of planet mass and disk viscosity. The
dotted contours, labeled with negative values, correspond to a positive torque and therefore \new{to} an outwards migration, in which case
the label indicates the semi-major axis doubling time. The dashed line is the \new{set} of  positions where the unperturbed viscous drift
rate $-3\nu/2r$ has the same absolute value as the inferred migration rate. Therefore the migration rate estimates are reliable only
\new{to} the right of the dashed line, i.e. wherever the material flow rate across the orbit is mostly accounted
for by the viscous drift. \new{The solid thick line corresponds to a vanishing total torque (limit of migration reversal).
The contours close to this limit correspond to migration times which tend to infinity, and they have been blanked for
numerical precision reasons.}}
\end{figure}

\begin{figure}
   \includegraphics[width=\columnwidth]{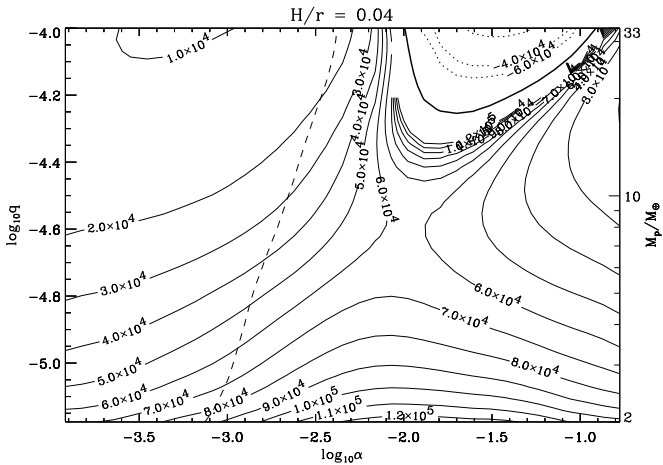}
\caption{\label{fig:taumig4}Migration time in years at~1~AU for a $h=0.04$ disk, as a function of planet mass and disk viscosity.
The contour line style obeys the same conventions as in Fig.~\ref{fig:taumig3}.}
\end{figure}

The migration regime corresponding to the range of masses, aspect ratios and viscosities investigated in this work corresponds
mostly to type~I migration regime, except the low aspect ratio, low viscosity and large mass runs, in which quite a significant gap 
is opened in the co-orbital region. As type~I migration is known to be too fast (i.e. the migration time of a protoplanetary core is
shorter than its build-up time), it is of interest to investigate the migration time from these results, as a function of aspect
ratio, planet mass and viscosity. Figs.~\ref{fig:taumig3} and~\ref{fig:taumig4} show the migration time, in years, of a
protocore at $1$~AU embedded respectively in a $3$\% and $4$\% aspect ratio disk. In either case the disk surface density is chosen to
be that of the minimum mass solar nebula ($1700$~g.cm$^{-2}$, i.e. $\Sigma_0=1.9\cdot 10^{-4}$, see Hayashi et al.~1985).
As the results of the high resolution runs of section~\ref{sec:highres} dealt only with  restricted mass and viscosity ranges, the results
shown at Figs.~\ref{fig:taumig3} and~\ref{fig:taumig4} correspond to the low resolution runs (i.e. to Figs.~\ref{fig:h03} and~\ref{fig:h04}),
for which the parameter coverage is much larger. As the results for the $h=0.04$ case in the high and low resolution case differ
sensibly, the contours of Fig.~\ref{fig:taumig4}  (and of Fig.~\ref{fig:taumig3}) 
should not be taken too literally. They do however have the merit to show the
behavior of the migration time as a function of mass and viscosity (even if the boundary of the domain of torque reversal should
not be taken literally) and they also have the merit to give a correct order of magnitude of this migration time, defined as:
\begin{equation}
\tau_\mathrm{mig}=\frac{M_pr_p^2\Omega_p}{2\Gamma}
\end{equation}
This migration time has been evaluated at $r_p=1$~AU, where the protoplanetary disk is likely to be subject
to the magneto-rotational instability (Balbus \& Hawley 1991), which endows it with a source of significant viscosity, which 
could be in the range of the values of $\alpha$ for which migration can reverse. Although this migration time is a few times
larger than previous results (e.g. Miyoshi et al. 1999 and references therein), it is still much shorter than the disk lifetime and
the core build-up time. It is therefore very unlikely that a unique protocore reaches the torque reversal mass before having
migrated all the way to the central object. One reason is that the torque reversal occurs in very thin disks, for which the
absolute value of the differential Lindblad torque is large (it scales as $h^{-2}$), and the associated migration time-scale is
short.
\new{One can see that the most favorable situation in both cases is for $\alpha=10^{-2}$ (i.e. for the corresponding viscosity
a given mass protocore has the largest migration time). For this value of $\alpha$ one gets bottleneck values of the
migration time respectively $7-8\cdot 10^4$~years (for $h=3$~\%) and $6-7\cdot 10^4$~years (for $h=4$~\%).}

It should also be noted that, as the corotation torque scales with the slope of the specific vorticity gradient, the torque reversal
that was found here for very thin disks is not likely to subsist if one takes a significant negative surface density gradient (i.e.
if $\Sigma \propto r^{-q}$ with $q\sim 1$).

 On the other hand if the specific vorticity gradient is even steeper than the one envisaged here, then the corotation torque, depending
on how steep this gradient is, may well unconditionally dominate the differential Lindblad torque. This might be the case
in the very central parts of the disk, where the outer disk solution has to be connected to an inner cavity (e.g. magnetically
cleared). If the transition region is larger than the co-orbital zone of an infalling body, then the corotation torque acting on it might be able
to counteract the differential Lindblad torque, providing another way of stopping the inward migrating bodies, as the disk
there is very likely to be viscous and therefore very likely to  be able to  sustain an unsaturated corotation torque.

As the analysis presented here is restricted to a planet held on a fixed circular orbit, it should be realized that the migration time-scale
estimated above is only valid whenever the viscous drift rate of material across the orbit is much larger than the planet inferred
drift rate $2\Gamma/(M_pr_p\Omega_p)$, otherwise the corotation torque expression has to explicitly include a term
proportional to $\dot a$, the migration rate, with a delay which scales as the outermost horseshoe libration
time. The domain for which the migration rate is negligible compared to the viscous drift radial velocity is shown in Figs.~\ref{fig:taumig3}
and~\ref{fig:taumig4}. 
\new{Therefore one can see that at very low viscosities [$\alpha < (1-3)\cdot 10^{-3}$] the expression for the corotation 
torque needs to be reconsidered, as most of the specific vorticity drift across the co-orbital region is not accounted for
by the viscous accretion but by the migration itself. In these circumstances the total torque felt by the planet depends
on its migration rate. The analyze and consequences of this feed-back will be presented elsewhere.}

\section{Summary}
This paper examines the torque exerted on a protoplanet held on a fixed circular orbit by a uniform surface density and uniform aspect
ratio viscous protoplanetary disk, by means of numerical simulations. A number of runs have been performed, varying the disk
aspect ratio, the planet mass and the disk viscosity, which enables one to disentangle the 
differential Lindblad torque/corotation torque additional term  from the co-orbital corotation
torque main term. The differential torque functional dependence upon the width of the librating fluid elements region and upon  viscosity
is in good agreement with previous expectations. 
The behavior of the total torque as a function of viscosity and planet mass is presented.
It is found that the curves of the normalized total torque as a function of viscosity are lift up as the planet
mass increases, eventually leading to a positive torque peak value in the thinnest disks. This behavior is explained as due
to the non-linear Lindblad torque cut-off and to the corotation torque additional term, which both conspire in lifting the
normalized torque as the planet mass increases. Smoothing issues are discussed. The smoothing length is shown to have
a different impact on the Lindblad torque and on the corotation torque, and the corotation torque is found 
to depend dramatically
on the smoothing length value, which shows that there is no such 
thing as a ``magic value'' for the smoothing length which would
give unconditionally correct results (i.e. compatible with fully three dimensional calculations).
A method is given to evaluate
 the value of the smoothing length which gives correct results for the corotation torque 
(i.e. identical to what one expects in a 3D situation). If the motion in the disk in the co-orbital region is purely horizontal, this method
should give in principle an exact result. Note however that the functional dependence upon viscosity 
of the 3D corotation torque and the 2D one with
the appropriate smoothing length may be different, and that the method exposed here is exact only for the case of a fully
unsaturated torque, i.e. at large viscosity (although before the cut-off).  
High resolution runs are presented in which the smoothing length
is chosen conservatively, and these runs show that Neptune-sized planets
undergo a positive torque in thin, viscous disks \new{with a shallow surface density profile}, 
although only a lower limit  of the torque value can be inferred from the runs.

\appendix

\section{Finite resolution effects on the co-orbital corotation torque estimate}
\label{ap:corot}
\new{
As the radial width of the co-orbital region in the simulations presented in this work amounts to a small number of
zone widths (typically~3 to~18), 
it is of interest to investigate how bad or good is the code used here in giving an estimate of the
co-orbital corotation torque. For this purpose, one can consider a simplified situation, such as the one depicted at
Fig.~\ref{fig:resol1}. This situation neglects the grid curvature and describes the gas flow around the planet in
a shearing sheet. In this situation, the corotation torque is given by
\begin{equation}
\label{eq:analyticshsh}
\Gamma_{ss}=\frac 34\Omega^2x_s^4r_p\frac{\partial\Sigma}{\partial r}
\end{equation}
\begin{figure}
   \includegraphics[width=\columnwidth]{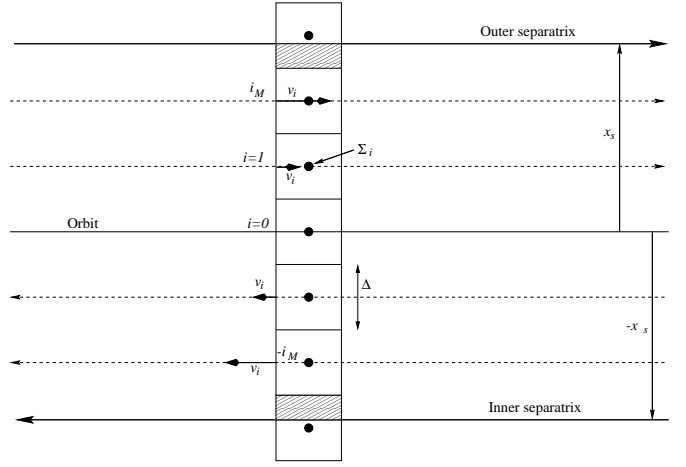}
\caption{\label{fig:resol1}
Sketch of a shearing sheet case. A radial set of zones is represented. The
separatrices are assumed to be horizontal and are symmetric w.r.t the orbit.
The azimuthal velocity is set on a staggered mesh (centered in radius,
staggered in azimuth), and is assumed in this simplified example to be
independent of azimuth.}
\end{figure}
The code used here enforces angular momentum conservation (to the computer accuracy) hence the balance of angular
momentum gained/lost by the orbit crossing fluid elements on horseshoe streamlines can be used as an estimate of the 
corotation torque. The angular momentum is a zone centered variable. In this simplified situation the velocity field is
assumed to be the unperturbed one. The zone radial index is denoted $i$, and vanishes at the orbit (as stated in 
section~\ref{sec:numeraspects}, the planet lies in the center of a zone). The index of the outermost zone fully embedded in
the co-orbital region is denoted~$i_M$. It is assumed that the co-orbital region is wide enough so that such a zone exists. The
expression for $i_M$ is therefore:
\begin{equation}
\label{eq:im}
i_M=E\left(\frac{x_s}{\Delta}-\frac 12\right)
\end{equation}
where $E(X)$ stands for the integer part of~$X$, and where $\Delta$ is the radial zone width.
The co-orbital corotation torque in this situation can therefore be expressed as:
\begin{equation}
\label{eq:ccnum1}
\Gamma^{\rm{num}}_{ss}=\Gamma^{\rm{num}^+}_{ss}+\Gamma^{\rm{num}^-}_{ss}
\end{equation}
where:
\begin{eqnarray}
\label{eq:ccnum2}
\Gamma^{\rm{num}^+}_{ss}&=&\sum_{i=0}^{i_M}\frac 32\Omega_p i\Delta^2\cdot \Omega_pr_p\Delta i\Sigma_i \\
&&+\left(\frac{x_s}{\Delta}-i_M-\frac 12\right)\frac 32\Omega_p^2 (i_M+1)^2\Delta^3r_p\Sigma_{i_M+1} \nonumber
\end{eqnarray}
is the torque due to the horseshoe fluid elements flowing from $i\Delta>0$ to $-i\Delta$, 
and where:
\begin{eqnarray}
\label{eq:ccnum3}
\Gamma^{\rm{num}^-}_{ss}&=&-\sum_{i=0}^{-i_M}\frac 32\Omega_p i\Delta^2\cdot \Omega_pr_p\Delta i\Sigma_i \\
&&-\left(\frac{x_s}{\Delta}-i_M-\frac 12\right)\frac 32\Omega_p^2 (i_M+1)^2\Delta^3r_p\Sigma_{-i_M-1} \nonumber
\end{eqnarray}
is the torque due to the horseshoe fluid elements flowing from $i\Delta<0$ to $-i\Delta$.
If one writes:
\begin{equation}
\label{eq:devsig}
\Sigma_i=\Sigma_0+i\Delta\frac{\partial\Sigma}{\partial r}+O[(i\Delta/r_p)^2]
\end{equation}
then combining Eqs.~(\ref{eq:ccnum2}) and~(\ref{eq:ccnum3}) one is led to:

\begin{eqnarray}
\label{eq:ccnum4}
\Gamma^{\rm{num}}_{ss}&\simeq& 3\Omega_p^2\Delta^3r_p\frac{\partial\Sigma}{\partial r}\\
&&\times\left[\sum_{i=0}^{i_M}i^3+(i_M+1)^3\left(\frac{x_s}{\Delta}-i_M-\frac 12\right)\right]\nonumber
\end{eqnarray}
The relative error on the corotation torque evaluation, ${\cal E}=(\Gamma^{\rm{num}}_{ss}-\Gamma_{ss})/\Gamma_{ss}$,
is therefore given by:
\begin{equation}
\label{eq:err1}
{\cal E}(Y) = \frac{4}{Y^4}\left[\sum_{i=0}^{i_M}i^3+(i_M+1)^3\left(Y-i_M-\frac 12\right)\right]
\end{equation}
where $Y=x_s/\Delta$ is the horseshoe region half-width, expressed in zone radial widths. The graph of the function
${\cal E}(Y)$ is the solid line of Fig.~\ref{fig:relerr}.

The second step of this error estimate consists of evaluating the discrepancy between the actual and effective 
 separatrix position. Indeed, the brackets of the second line of Eqs.~(\ref{eq:ccnum2})
and~(\ref{eq:ccnum3}) stand for the mass fraction of the outermost horseshoe zone included within the separatrix (i.e. it is
 also the ratio of the shaded area surface to the zone 
surface in Fig.~\ref{fig:resol1}). There is no warranty however that the separatrix position which has to be used in this
evaluation coincides with the one provided by the streamline analysis. In other words, one has to evaluate how much
separatrix crossing numerical effects can lead to, in order to infer how distant the actual and effective separatrices can be.
One approximate way to do that is to consider a 1D problem (along the radial dimension). Disregarding azimuthal advection
should not lead to significant errors, as this latter is treated apart during the advection procedure based on the
operator splitting technique, and as azimuthal motion  should not lead to separatrix crossing provided this latter is
horizontal (or weakly tilted) in the $(\theta, r-r_p)$ plane.
\begin{figure}
   \includegraphics[width=\columnwidth]{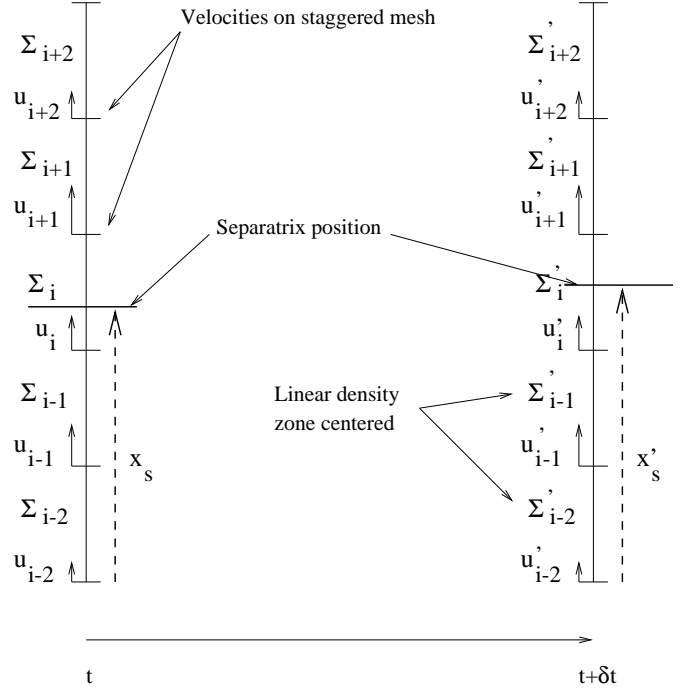}
\caption{\label{fig:resol2}
Sketch of the 1D mesh used to evaluate the impact of numerical effects on
separatrix crossing. The linear density $\Sigma$ is zone-centered, while
the radial velocity $u$ is defined at zone interfaces (on a staggered mesh).}
\end{figure}
Following the notations of Fig.~\ref{fig:resol2}, one can write the total mass $M_-$ contained below the separatrix (or 
below any arbitrary position $x_s$) at time $t$ as:
\begin{equation}
M_- = \Delta\sum_{j=0}^{i-1}\Sigma_j+\Sigma_i(x_s-i\Delta)
\end{equation}
where $\Delta$ is the zone spacing. This mass will be referred to hereafter as the
inner mass. At time $t+\delta t$, the inner mass becomes:
\begin{eqnarray}
M_-' &=& \Delta\sum_{j=0}^{i-1}\Sigma_j'+\Sigma_i'(x_s'-i\Delta)
\end{eqnarray}
where the primes denote the new quantities, and where one has the following relationship:
\begin{equation}
\Sigma_j'=\Sigma_j+\frac{\delta t}{\Delta}(F_j-F_{j+1})
\end{equation}
where $F_j$ is the mass flux (oriented upwards) across the interface between the zones $j-1$ and $j$. One also needs the
following relationship:
\begin{equation}
x_s'=x_s+\delta t\left[u_i\left(i+1-\frac{x_s}{\Delta}\right)+u_{i+1}\left(\frac{x_s}{\Delta}-i\right)\right]
\end{equation}
which comes from the fact that the streamlines (and therefore the separatrix) are found integrating the linearly interpolated
velocity field.
One can therefore write:
\begin{eqnarray}
M_-'&=&\Delta\sum_{j=0}^{i-1}\Sigma_j+\delta t\sum_{j=0}^{i-1}(F_j-F_{j+1})\nonumber \\
&&+\left[\Sigma_i+\frac{\delta t}{\Delta}(F_i-F_{i+1})\right](x_s'-i\Delta)
\end{eqnarray}
The mass fluxes can be written as:
\begin{equation}
F_j=\Sigma_j^*u_j
\end{equation}
where $\Sigma_j^*$ is an estimate of the linear density at the interface, the expression of which depends on the
numerical method (Stone \& Norman 1992). To lowest order in $u_j\delta t/\Delta$, the mass variation can be expressed, after
a few transformations,  and assuming $F_0=0$, as:
\begin{eqnarray}
\delta M_-&=&\delta t\left[u_i(\Sigma_i-\Sigma_i^*)\left(i+1-\frac{x_s}{\Delta}\right)\right.\nonumber \\
&&\left.+u_{i+1}(\Sigma_i-\Sigma^*_{i+1})\left(\frac{x_s}{\Delta}-i\right)\right]
\end{eqnarray}
\begin{figure}
   \includegraphics[width=\columnwidth]{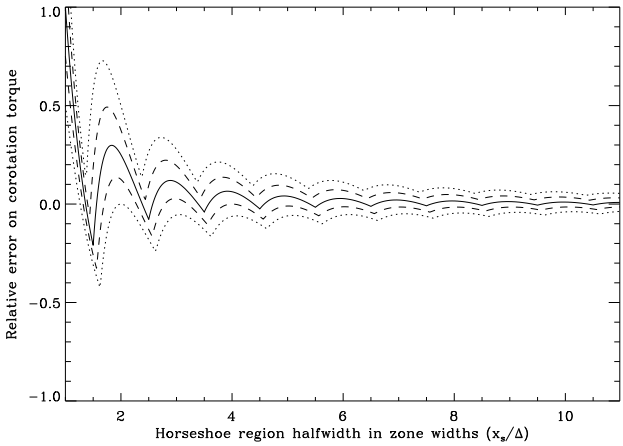}
\caption{\label{fig:relerr}Relative error on the co-orbital corotation torque
in the shearing sheet, as a function of resolution (number of zones in
the horseshoe region half-width). The dotted and dashed lines are respectively
for the worst case and a conservative case for the runs presented here
(see text for details).}
\end{figure}
This mass variation necessarily corresponds to the separatrix crossing mass, since $F_0=0$.
One can get an estimate of this mass variation assuming $u_i=u_{i+1}\equiv u$ (a reasonable assumption in the case of a radial
viscous drift) and writing $\Sigma_i-\Sigma^*_i=\Sigma_{i+1}^*-\Sigma_i\simeq\partial_r\Sigma\Delta/2$ (in which case one
has to assume that $|u\delta t|\ll\Delta$, a reasonable assumption for the case of interest here).
One is then led to:
\begin{equation}
\delta M_-\simeq u\,\delta t\,\partial_r\Sigma(x_i^c-x_s)
\end{equation}
where $x_i^c=(i+1/2)\Delta$ is the $x$ value of the zone center. One can therefore write:
\begin{equation}
\frac{dM_-}{dt} =  u(x_i^c-x_s)\partial_r\Sigma  = \dot{x_s}(x_i^c-x_s)\partial_r\Sigma
\end{equation}
from which one can infer:
\begin{equation}
M_-=M_-^c-\frac 12(x_c^i-x_s)^2\partial_r\Sigma
\label{eq:massin}
\end{equation}
where $M_-^c$ is the inner mass when the separatrix crosses the zone center. Eq.~(\ref{eq:massin}) shows that the inner mass
at the top ($x_s=x_c^i+\Delta/2$) and the bottom ($x_s=x_c^i-\Delta/2$) of the zone are the same, from which one
can conclude that, although some
mass can cross the separatrix as its sweeps the zone, there is no net effect as it sweeps the entire zone from one interface
to the other. As a consequence,
the maximum amount of mass that can cross the separatrix as it sweeps an arbitrary number of zones is 
$M_{sc}=\partial_r\Sigma\cdot\Delta^2/8$.
Denoting with $\delta x_s$ the error on the separatrix position, which is defined as: $\delta x_s \cdot\Sigma=M_{sc}$, one is led to
the following relation:
\begin{equation}
\delta x_s = \frac{\Delta^2}{8}\frac{\partial_r\Sigma}{\Sigma}
\end{equation}
The worst case is obtained for $\partial_r\Sigma\sim\Sigma/\Delta$, i.e. when the separatrix lies on the edge of a fully emptied gap,
and when this edge is not resolved (i.e. when the density falls abruptly from $\Sigma$ to $0$ from one zone to its neighbor). It
should be stressed that this situation is far from being met in the simulations presented here. Even in this worst case, the
maximal error on the separatrix position will be: $\delta x_s=\Delta/8$, whereas in a more reasonable case for which $\partial_r\Sigma$
will typically amount to a small fraction of $\Sigma/H$, the maximal error on the separatrix position will be a fraction of 
$(\Delta/H)\cdot \Delta/8$. The worst case is displayed in Fig.~\ref{fig:relerr} with dotted lines, while the conservative case
where $\delta x_s=\Delta /16$, obtained for $\partial_r\Sigma=\Sigma/H$ and $H=0.03$ in the normal resolution runs, is displayed
with dashed lines. It is worth noting that the separatrix error is much smaller than this conservative estimate, in particular for
small planet masses, which hardly perturb the disk density profile, and for which $Y=x_s/\Delta$ is the smallest. One can conclude
from Fig.~\ref{fig:relerr} that the corotation torque 
is described with an accuracy better than $15$~\% as soon as $x_s/\Delta\simgeq 2.3$.
Although this condition is not met for the smallest mass planets  in the normal resolution runs presented here, the corotation
torque, which on average for $1\leq Y\leq 2$ numerical effects tend to overestimate, is never found to be large enough to
cancel out the differential Lindblad torque for these objects. Therefore, one can conclude that even with a relatively small
number of mesh rings involved in the co-orbital region, the code used here can describe with a reasonable precision the dynamics
of the co-orbital corotation torque, and that whenever the planet mass is too small for the dynamics of this region to be
correctly described (for the resolution and parameter set used here), 
the corotation torque is too small to significantly affect the migration anyway (one recovers the linear
regime, see section~\ref{sec:intro} for details).

}
\begin{acknowledgements}
I wish to thank Prof. M. Tagger for fruitful discussions, as well as the anonymous referee for comments and a careful reading
of the manuscript.
The early stages of this work were supported by the research network ``Accretion onto Black Holes,
Compact Stars, and Protostars'' funded by the European Commission under contract ERBFMRX CT 98-0195.
Computational resources were available at the Rechenzentrum Garching and at the CGCV Grenoble, and are
gratefully acknowledged. The last stages of this work were  carried out at the Institute of Astronomy, UNAM, Mexico,
and I wish to thank Drs. J. Cant\'o,  A. Raga and A. Santill\'an for hospitality.
\end{acknowledgements}

\end{document}